\documentstyle[aps,eqsecnum]{revtex}

\def\simless{\mathbin{\lower 3pt\hbox
   {$\rlap{\raise 5pt\hbox{$\char'074$}}\mathchar"7218$}}} 
\def\simgreat{\mathbin{\lower 3pt\hbox
   {$\rlap{\raise 5pt\hbox{$\char'076$}}\mathchar"7218$}}} 
\def\go{\mathrel{\raise.3ex\hbox{$>$}\mkern-14mu\lower0.6ex\hbox{$\sim$}}}
\def\lo{\mathrel{\raise.3ex\hbox{$<$}\mkern-14mu\lower0.6ex\hbox{$\sim$}}}
\def\be{\begin{equation}}
\def\ee{\end{equation}}

\def\sc{{{\rm sc}}}
\def\eps{{\epsilon}}
\def\bk{{\bf k}}
\def\bo{{\bf \Omega}}
\def\bp{{\bf p}}
\newcommand{\calp}{ {\cal P} }
\newcommand{\beqa}{\begin{eqnarray}}
\newcommand{\eeqa}{\end{eqnarray}}
\newcommand{\overl}{\overline}
\newcommand{\cv}{c_V}
\newcommand{\ca}{c_A} 
\newcommand{\bbhat}{\hat{\bf B}}
\newcommand{\bhat}{\hat{B}}
\newcommand{\gev}{\rm GeV}

\newcommand{\Kslash}{\not{\! K}}
\newcommand{\tr}{\rm Tr}
\newcommand{\kt}{T}
\newcommand{\bq}{{\bf q}}
\newcommand{\fnu}{f_{\nu}}
\newcommand{\fN}{f_N}
\newcommand{\bh}{{\bf h}}

\newcommand{\grad}{{\bf \nabla}}

\newcommand{\bx}{{\bf x}}

\newcommand{\kabs}{\kappa^{({\rm abs})}}
\newcommand{\kabsstar}{\kappa^{*({\rm abs})}}
\newcommand{\kscat}{\kappa^{(\rm sc)}}
\newcommand{\ktot}{\kappa^{(\rm tot)}}

\newcommand{\cH}{\cal H}
\newcommand{\cT}{{\cal T}}

\begin{document}
\tightenlines
\draft

\title{ Neutrino-Nucleon Interactions in Magnetized Neutron-Star Matter:
The Effects of Parity Violation }

\author{Phil Arras~\footnote{Department of Physics, Cornell University;
E-mail: arras@spacenet.tn.cornell.edu}
and  Dong Lai~\footnote{Department of Astronomy, Cornell University;
dong@spacenet.tn.cornell.edu}}
\address{Center for Radiophysics and Space Research, Space Sciences
Building\\
Cornell University, Ithaca, NY 14853}

\date{\today} 
\maketitle

 

\begin{abstract}
We study neutrino-nucleon scattering and absorption in
a dense, magnetized nuclear medium. These are the most important sources 
of neutrino opacity governing the cooling of a proto-neutron star
in the first tens of seconds after its formation. Because the
weak interaction is parity violating, the absorption and scattering 
cross-sections depend asymmetrically on the directions of the neutrino 
momenta with respect to the magnetic field. We develop the moment 
formalism of neutrino transport in the presence of such asymmetric 
opacities and derive explicit expressions for the neutrino flux and 
other angular moments of the Boltzmann transport equation. 
For a given neutrino species, there is a drift flux of neutrinos along the
magnetic field in addition to the usual diffusive flux. This
drift flux depends on the deviation of the neutrino 
distribution function from thermal equilibrium.
Hence, despite the fact that the neutrino cross-sections are 
asymmetric throughout the star, asymmetric neutrino flux can be generated 
only in the outer region of the proto-neutron star where the neutrino 
distribution deviates significantly from thermal equilibrium.
The deviation from equilibrium is similarly altered by the asymmetric
scattering and absorption, although its magnitude will still be quite
small in the interior of the star. 
We clarify two reasons why previous studies have led to misleading 
results. First, inelasticity must be included in the phase space 
integrals in order to satisfy detail balance. Second, nucleon recoil 
must be included in order to find the leading order asymmetric
cross sections correctly, even though it can be ignored to leading order
to get the zero field opacities.
In addition to the asymmetric absorption opacity arising from nucleon
polarization, we find the contribution of the electron (or 
positron) ground state Landau level. For neutrinos of energy
less than a few times the temperature, this is the dominant source of
asymmetric opacity.
Lastly, we discuss the implication of our result to
the origin of pulsar kicks: in order to generate kick velocity of
a few hundred km~s$^{-1}$ from asymmetric neutrino emission 
using the parity violation effect, the proto-neutron star must have
a dipole magnetic field of at least $10^{15}-10^{16}$~G. 
\end{abstract}
\bigskip

 

\section{\bf INTRODUCTION}

\subsection{Astrophysical Motivation}

Neutrinos play an essential role in core-collapse supernovae
and the formation of neutron stars\cite{Bethe90,Petschek90}.
It is through neutrino emission that 
a hot proto-neutron star releases its gravitational binding energy and
cools. The explosion itself also relies 
on the neutrinos for its success. The most important 
ingredient of neutrino 
transport in proto-neutron stars is the neutrino opacity
in a dense nuclear medium. Much effort has been devoted to 
understanding various effects of neutrino-matter interactions at 
supra-nuclear density (e.g., 
\cite{Tubbs75,Sawyer79,Iwamoto82,Bruenn85,Horowitz92,Keil95,Sigl96,Reddy97,Burrows98} and references therein).
Neutron stars, however, possess strong 
magnetic fields. While the present-day, dipolar magnetic fields
of most radio pulsars lie in the range of $10^{12}-10^{13}$~G, 
it has been suggested that fields of $10^{15}$~G or larger
can be generated by dynamo process in proto-neutron stars\cite{Thompson93}.
Several recent observations\cite{Vasisht97,Kouveliotou98a,Kouveliotou98b}
have lent support to the idea that soft
gamma-ray repeaters and slowly spinning X-ray pulsars
(``anomalous x-ray pulsars") in supernova
remnants are neutron stars endowed with superstrong magnetic fields
$B\go 10^{14}$~G\cite{Thompson95,Thompson96}. 
It is therefore necessary to understand 
how neutrino opacities are modified by the presence of a strong
magnetic field. This is the subject of our paper.

A direct motivation of our study is to explore whether strong
magnetic fields can induce asymmetric neutrino emission from 
proto-neutron stars to explain pulsar ``kicks''.
It has long been recognized that neutron stars have 
space velocities that are about an order of 
magnitude greater than their progenitors' (e.g., \cite{Gunn70,Trimble68}). 
Recent studies of pulsar proper motion give $200-500$~km~s$^{-1}$
as the mean 3D velocity of neutron stars at birth
\cite{Lyne94,Lorimer97,Hansen97,Cordes97}, with possibly a
significant population having velocity of order or greater than 
$700$~km~s$^{-1}$. Direct evidence for pulsars with velocities 
$\go 1000$~km~s$^{-1}$ comes from observations of the bow shock 
produced by PSR B2224+65 in the interstellar 
medium\cite{Cordes93}, and studies of pulsar-supernova 
remnant associations\cite{Frail94}. A natural explanation for
such high velocities is that supernova explosions are
asymmetric, and provide kicks to nascent neutron stars. 
Support for supernova kicks has come from the detections of geodetic 
precession in the binary pulsar PSR 1913+16\cite{Cordes90,Arzoumanian96} 
and the orbital plane precession in the 
PSR J0045-7319/B star binary and its fast orbital 
decay\cite{Kaspi96,Lai96}. In addition, evolutionary studies
of the neutron star binary population imply the existence of pulsar
kicks\cite{Deway87,Brandt95,Fryer97,Fryer98}.

Two classes of mechanisms for the {\it natal kicks}
have been suggested. The first class relies on hydrodynamical 
instabilities in the collapsed stellar 
core\cite{Burrows92,Burrows95,Janka94,Janka96,Herant94} that lead to 
asymmetric matter ejection and/or asymmetric neutrino emission;
but numerical simulations indicate that these instabilities 
are not adequate to account for kick velocities 
$\go 100$~km~s$^{-1}$\cite{Janka94,Burrows96,Janka98}. 
Global asymmetric perturbations in the presupernova cores
are required in order to produce the observed 
kicks\cite{Burrows96,Goldreich96}.
In this paper we are concerned with the second class of models
in which the pulsar kicks arise from magnetic field induced 
asymmetry in neutrino emissions from proto-neutron stars.
The fractional asymmetry $\alpha$ in the radiated neutrino energy
required to generate a kick velocity $V_{\rm kick}$ is
\be
\alpha=0.028\,\left({M\over 1.4M_\odot}\right)\left({3\times 10^{53}~{\rm erg}
\over E_{\rm tot}}\right)\left({V_{\rm kick}\over 1000~{\rm km~s}^{-1}}\right),
\label{alpha}\ee
where $M$ is the mass of the neutron star and $E_{\rm tot}$ is the
total neutrino energy radiated from the neutron star.
Since $99\%$ of the neutron star binding energy (a few times $10^{53}$~erg)
is released in neutrinos, tapping the neutrino energy would appear to be
an efficient means to kick the newly-formed neutron star.
Magnetic fields are naturally invoked to break the spherical symmetry in
neutrino emission, but the actual mechanism is unclear.
We first review previous work related to this subject.

\subsection{Review of Previous Work}

Beta decay in a strong magnetic field was first investigated 
in Refs.~\cite{Matese69,Fassio-Canuto69} (See also 
Refs.~\cite{Lai92,Cheng93}).
A number of authors have noted that parity violation in weak
interactions may lead to asymmetric neutrino emission from
proto-neutron stars\cite{Chugai84,Dorofeev85,Vilenkin95,Horowitz97a}.
Chugai\cite{Chugai84} and Vilenkin\cite{Vilenkin95} (see also 
Ref.~\cite{Bezchastnov96}) considered neutrino-electron scattering and
concluded that the effect is extremely small\footnote{Note that Chugai's
estimate for the electron
polarization in the relativistic and degenerate regime (the relevant
physical regime) is incorrect. This error leads to an overestimate
of the effect as compared to Vilenkin's result.}
(e.g., to obtain $V_{\rm kick}=300$~km~s$^{-1}$ would require
a magnetic field of at least $10^{16}$~G). However, neutrino-electron
scattering is less important than neutrino-nucleon scattering in 
determining the characteristics of neutrino transport in proto-neutron 
stars (e.g., \cite{Burrows86,Prakash97}).
Similarly, Dorofeev et al.\cite{Dorofeev85} considered neutrino 
emission by Urca processes in strong magnetic fields. 
However, as we shall see below (see Ref.~\cite{Lai98a}),
the asymmetry in neutrino emission is cancelled by the
asymmetry associated with neutrino absorption for young proto-neutron
stars where the neutrinos are nearly in thermal equilibrium. 
The size of the asymmetric flux due to absorption/emission processes
is then dependent on the {\it deviations} from thermal equilibrium
at the neutrino photosphere.

Horowitz \& Li\cite{Horowitz97b} suggested that large asymmetries in
the neutrino flux could result from the {\it cumulative} effect of
multiple scatterings of neutrinos by nucleons 
which are slightly polarized by
the magnetic field. In particular, they found that the 
size of the asymmetry was proportional to the optical depth of the 
star to neutrinos ($\tau \sim 10^4$). The result was that 
kick velocities of a few hundred km~s$^{-1}$ could be generated by
field strengths of only a few times $10^{12}$~G. Initial neutrino 
cooling calculation\cite{Lai98a} of a proto-neutron star in magnetic fields
appeared to indicate that a dipole field of order
$10^{14}$~G is needed to produce a kick velocity of $200$~km~s$^{-1}$.
The larger magnetic field required results from cancellations
of the asymmetries associated with $\nu_\mu,~\nu_\tau$ and their 
antiparticles as well as the opposite signs of polarizations
of neutrons and protons. Preliminary numerical study reported in 
Ref.~\cite{Janka98} drew a similar conclusion although it was claimed 
that only $10^{13}$~G is needed to produce $200$~km~s$^{-1}$. 

As appealing as the cumulative effect may be, we now believe
that it does not work in the bulk interior of the 
neutron star\cite{arraslai98a,Kusenko98}, and the conclusions reached
in Refs.~\cite{Janka98,Lai98a,Horowitz97b} are incorrect \cite{credit}. 
In spite of the fact that the scattering cross-section
is asymmetric with respect to the magnetic field for individual neutrinos, 
detailed balance requires that no asymmetric neutrino flux can arise in 
the stellar interior where neutrinos are in thermal 
equilibrium to a good approximation.
Since this issue is somewhat subtle and counter-intuitive, 
we discuss it in detail in \S II where we also point out where 
previous studies went wrong.

A related, but different kick mechanism relies on the 
asymmetric magnetic field distribution in proto-neutron 
stars\cite{Bisnovatyi-Kogan93,Roulet97,Lai98b}. 
Since the cross section for $\nu_e$ ($\bar\nu_e$) absorption on neutrons
(protons) depends on the local magnetic field strength due to
the quantization of energy levels for the $e^-$ ($e^+$) 
produced in the final state, the local neutrino fluxes emerged from different
regions of the stellar surface are different. It was found\cite{Lai98b}
that to generate a kick velocity of $\sim 300$~km~s$^{-1}$ using this 
mechanism alone would require that the difference 
in the field strengths at the two opposite poles of the star 
be at least $10^{16}$~G. 

There have also been several interesting ideas on pulsar kicks
which rely on nonstandard neutrino physics. It was
suggested\cite{Kusenko96} that asymmetric $\nu_\tau$ emission could 
result from the Mikheyev-Smirnov-Wolfenstein flavor transformation
between $\nu_\tau$ and $\nu_e$ inside a magnetized proto-neutron star
because a magnetic field changes the resonance condition for the
flavor transformation. Another similar idea\cite{Akhmedov97}
relies on both the neutrino mass and the neutrino magnetic 
moment to facilitate the flavor transformation. 
More detailed analysis\cite{Qian97,Janka98b}, however, indicates
that even with favorable neutrino parameters (such as
mass and magnetic moment) for neutrino oscillation, the induced pulsar 
kick is much smaller than previously estimated. We will not consider
the issues related to nonstandard physics in this paper. 

Finally, we mention that previous calculations of neutrino processes
in magnetic fields have generally neglected nucleon recoils (e.g., 
\cite{Janka98,Matese69,Fassio-Canuto69,Dorofeev85,Lai98a,Horowitz97b}).
Although this simplification is justified in most cases 
without magnetic field, it is invalid in a magnetic field because the {\it
asymmetric part} of the opacity depends sensitively on the
phase space available in the scattering/absorption. This and other
technical issues
(such as Landau levels) will be addressed in our paper. 

\subsection{Plan of This Paper}

In this paper, we carry out a systematic study of
neutrino-nucleon ($\nu,N$) scattering and electron neutrino 
absorption/emission ($\nu_e+n \rightleftharpoons p+e^{-}$ and 
$\bar{\nu}_e+p \rightleftharpoons n+e^{+}$) in a dense,
magnetized nuclear medium. These are the most important sources of
opacity for neutrino cooling of the proto-neutron star
in the first tens of seconds after its formation, when most of
the binding energy of the neutron star is radiated as neutrinos. 
Our study goes beyond merely calculating differential cross-sections
of the neutrinos in magnetized medium in that we derive the expression
for the neutrino flux and other angular moments from Boltzmann equation. 
This is necessary in order to determine whether the effect of 
parity violation introduces any asymmetric ``drift flux''
in addition to the usual ``diffusive flux''. Indeed, there are a number of
subtleties in these derivations that several previous studies have arrived at
incorrect results (see \S I.B and \S II).
We show that, despite of the fact that the scattering/absorption 
cross-sections are asymmetric with respect to the magnetic field
for individual neutrinos, there is no ``drift flux'' when the
neutrinos are in thermal equilibrium; the drift flux is proportional to the
{\it deviations} from equilibrium, which are small below the
neutrino-matter decoupling layer. Hence, asymmetric neutrino emission
can be generated only near the surface layer of the star. 

In \S II we discuss a simplified calculation and point out its problems. 
This serves as an illustration of various issues that one must pay attention 
to in order to obtain the correct answers. We begin our 
formal theoretical development in \S III, where the relevant
cross sections are defined starting from the Boltzmann transport equation. 
It is important to distinguish different cross sections (e.g.,
those related to scattering into the beam and scattering out of the beam
in the Boltzmann equation) in order to satisfy the principle of detailed
balance, which states that in complete thermal equilibrium the
collisional term in the Boltzmann equation vanishes. 

Section IV contains a detailed calculation of $\nu-N$ 
scattering in magnetic fields. Starting from the weak interaction 
Hamiltonian, we compute the scattering opacity, carefully including the 
effect of nuclear motion to lowest nonvanishing order. 
This opacity is then used to find the contribution to the 
angular moments of the transport equation. Explicit expressions 
are obtained for the outer layer of the neutron star where
nucleons are nondegenerate and where parity-violating
asymmetric flux can be generated.
A technical complication arises from 
the effect of small inelasticity: Even in the regime where the nucleon 
recoil energy is much smaller than the neutrino energy and temperature,
phase space considerations require that the inelasticity effect be included
when deriving the asymmetric flux and other moments (The situation is  
similar to that found in the derivation of the Kompaneets equation in
electron-photon scattering; see Ref.\cite{pomraning73}).  

In Section V we calculate the cross sections for the absorption of
$\nu_e$ and $\bar\nu_e$ by nucleons. As in the scattering case,
it is necessary to include nucleon recoil in the absorption calculation.
Additional complications arise from the quantized Landau levels of the final
state electrons (or positrons). 
In particular, the ground state Landau level of the electron 
introduces an effective electron ``polarization'' 
term in the asymmetric part
of the opacity. For certain parameter regimes, 
this electron ``polarization''
term dominates the asymmetry in the absorption opacity. 
We also demonstrate explicitly that the Landau levels of protons
have no effect on the absorption opacity since many levels are summed
over for the situation of interest.

In Section VI we combine the results of \S IV and \S V to derive
the angular moments of the Boltzmann transport equation. These moment 
equations are truncated at the quadrupole order, since we expect 
that the contributions of the higher order terms to the asymmetric flux are
smaller. As expected, our explicit expression for the neutrino flux contains the
usual diffusive flux plus a drift flux along the magnetic field. This drift
flux, however, depends on the deviation for the neutrino distribution from
thermal equilibrium. Although strictly speaking our truncated moment equations 
break down near neutron star surface, these equations are accurate
below the neutrinosphere, and provide a reasonable physical description of
the neutrino radiation field throughout the star. Finally in \S VII we
use the moment equations to obtain an order-of-magnitude estimate of 
the asymmetric neutrino emission from the proto-neutron star due to 
the parity violating processes considered in this paper.  

Throughout this paper, we treat nucleons as noninteracting particles 
This is clearly a simplifying assumption. In reality, 
strong interaction correlations may significantly change the
neutrino opacities (e.g., \cite{Reddy97,Burrows98} and references therein).
However, the goal of this paper is to consider whether there is
any new effect associated with strong magnetic fields. For this purpose,
it is certainly appropriate to focus on noninteracting nucleons, 
particularly since there are still large uncertainties in our 
understanding of matter at super-nuclear densities. Moreover, 
for certain nuclear potentials, the nuclear medium effects merely amount
to giving the nucleon an effective mass, and therefore the result
of this paper can be easily extended. For general nuclear interactions,
it is likely that the qualitative conclusion reached in this paper 
will remain valid, although this issue is beyond the scope of this paper. 
 
Unless noted otherwise, we shall use units in which $\hbar,~c$ and the
Boltzmann constant $k_B$ are unity.

\section{A Simplified Calculation of the Scattering Effect and 
its Problems}

As indicated above, the effect of asymmetric neutrino-nucleon scattering 
is sufficiently subtle that several previous studies have led to 
a wrong conclusion. It is therefore instructive to consider 
a simplified treatment of the problem to understand how
previous work went wrong. This section also serves as an illustration
of the various issues that one must pay attention to in 
doing such a calculation
(Our systematic calculation is presented in \S\S III-VI). 

\subsection{The Calculation}

We shall follow the treatment as given in Ref.~\cite{Lai98a} (Refs.
\cite{Janka98,Horowitz97b} used a Monte-Carlo method for the neutrino
transport, which is less transparent for our analysis).
For the purpose of clarity, we consider the scattering of
neutrinos by nondegenerate neutrons. Assuming the scattering is 
elastic (which is a good approximation since the neutrino energy $k$
is much less than the neutron mass), one can easily obtain 
the matrix elements when the initial neutron has spin along the
magnetic axis (the $z$-axis):
\begin{eqnarray}
\left|\cH_{\bk\uparrow\rightarrow\bk'\uparrow}\right|^2
&=&2\,G_F^2c_V^2\,\left[\cos{1\over 2}(\theta'-\theta)
+\lambda\cos{1\over 2}(\theta+\theta')\right]^2,\label{elem1}\\
\left|\cH_{\bk\uparrow\rightarrow\bk'\downarrow}\right|^2
&=&2\,G_F^2c_A^2\,\left(2\sin{1\over2}\theta'\cos{1\over 2}\theta\right)^2,
\label{elem2}
\end{eqnarray}
where $\bk$ ($\bk'$) is the initial (final) neutrino momentum, 
$\theta$ ($\theta'$) is the angle between $\bk$ ($\bk'$) and the $z$-axis 
(assuming azimuthal angle $\phi=0$), and $G_F,~c_V,~c_A,~\lambda$ are 
weak interaction constants as defined in Appendix A. 
The differential cross section,
for the initial nucleon with spin along the $z$-axis, is given by
\begin{eqnarray}
\left({d\sigma\over d\Omega'}\right)_{\bk\uparrow\rightarrow\bk'}
&=&{k^2\over(2\pi)^2}\left[
\left|\cH_{\bk\uparrow\rightarrow\bk'\uparrow}\right|^2
+\left|\cH_{\bk\uparrow\rightarrow\bk'\downarrow}\right|^2\right]\nonumber\\
&=&{k^2\over (2\pi)^2}G_F^2c_V^2\biggl[(1+3\lambda^2)
+2\lambda(\lambda+1)\cos\theta-2\lambda(\lambda-1)\cos\theta'\nonumber\\
&&~~~~~~~~~~+(1-\lambda^2)\cos(\theta-\theta')\biggr].
\end{eqnarray}
One can similarly obtain $(d\sigma/d\Omega')_{\bk\downarrow\rightarrow\bk'}$
when the initial nucleon spin is anti-parallel to the $z$-axis.
For general nucleon spin polarization $P=\langle\sigma_z\rangle$,
the differential cross section is given by
\begin{eqnarray}
\left({d\sigma\over d\Omega'}\right)_{\bk\rightarrow\bk'}
&=&\left({1+P\over 2}\right)\left({d\sigma\over 
d\Omega'}\right)_{\bk\uparrow\rightarrow\bk'}
+\left({1-P\over 2}\right)
\left({d\sigma\over d\Omega'}\right)_{\bk\downarrow\rightarrow\bk'}
\nonumber\\
&=&\left({G_Fc_Vk\over 2\pi}\right)^2(1+3\lambda^2)
\biggl[1+\epsilon_{\rm in}\cos\theta+\epsilon_{\rm out}\cos\theta'
+{1-\lambda^2\over 1+3\lambda^2}\cos(\theta-\theta')\biggr],
\label{scatt2}
\end{eqnarray}
where 
\be
\epsilon_{\rm in}=2P{\lambda(\lambda+1)\over(1+3\lambda^2)},~~~~
\epsilon_{\rm out}=-2P{\lambda(\lambda-1)\over(1+3\lambda^2)}.
\ee
This clearly indicates that the scattering is asymmetric with respect
to the magnetic field, a direct consequence of parity violation 
in weak interaction. A similar expression was derived in
Ref.~\cite{Horowitz97b}, although a different sign in
$\epsilon_{\rm in}$ and $\epsilon_{\rm out}$ was given. 

Next we study the consequence of the asymmetric cross section 
on the neutrino flux. The Boltzman transport equation for the neutrino
distribution function $\fnu(\bk)$ can be written in the form:
\beqa
\frac{\partial \fnu(\bk)}{\partial t} + \bo \cdot \nabla \fnu(\bk)
=&&
\int\!d\Omega'\left({d\kappa\over d\Omega}\right)_{\bo'\rightarrow\bo}
\!\!f_\nu(\bk')\left[1-f_\nu(\bk)\right]\nonumber\\
&&-\int\!d\Omega'\left({d\kappa\over d\Omega'}\right)_{\bo\rightarrow\bo'}
\!\!f_\nu(\bk)\left[1-f_\nu(\bk')\right],
\label{scatt}
\eeqa
where $\bo$ and $\bo'$ are unit vectors along $\bk$ and $\bk'$, 
respectively, and the (elastic) differential cross section per unit volume
can be written in the form
\beqa
\left({d\kappa\over d\Omega'}\right)_{\bo\rightarrow\bo'}
&&={\kappa\over 4\pi}\Bigl[1+\epsilon_{\rm in}\bo\cdot\bbhat
+\epsilon_{\rm out}\bo'\cdot\bbhat
\nonumber \\ & + &
{\rm const.} (\bo\cdot\bo')\Bigr],
\label{elasticcross}
\eeqa
with $\bbhat$ the unit vector along the magnetic field. 
Note that in eq.~(\ref{scatt}) we have neglected neutrino absorption 
for simplicity. The first order moment of the transport equation 
is obtained by multiplying eq.~(\ref{scatt}) by $\bo$ and then integrating
over $d\Omega$.
The specific neutrino flux is then given by\footnote{
An overall factor of $(1-{\rm const.}/3)^{-1}$ has been dropped in this
equation.}
\be
{\bf F}_\nu=-{c\over 3\kappa}\nabla U_\nu+{1\over 3}(\epsilon_{\rm out}
-\epsilon_{\rm in})c U_\nu (1-f_\nu)\bbhat,
\label{fluxx}
\ee
where $U_\nu$ is the specific neutrino energy density. 
According to eq.~(\ref{fluxx}), the neutrino flux consists of the usual
diffusive flux, ${\bf F}_{\rm diff}\propto \nabla U_\nu$, and a ``drift'' 
flux ${\bf F}_{\rm drift}$ along the 
magnetic field. The drift flux induces asymmetric neutrino transport. 
One can easily see that the ratio $F_{\rm diff}/F_{\rm drift}$ is of order
$(\eps_{\rm out}-\eps_{\rm in})\tau$, where $\tau\sim \kappa R$ 
is the optical depth of the star ($R$ is the stellar radius),
i.e., the asymmetry increases with $\tau$. This is the origin of the
cumulative effect discussed in 
Refs.~\cite{Janka98,Lai98a,Horowitz97b}.
 
\subsection{The Problems}

The calculation presented above, while physically motivated, is actually
incorrect. There are two problems:
first, the asymmetry terms (those proportional to 
$\eps_{\rm in}$ and $\eps_{\rm out}$) in eq.~(\ref{scatt2}) are
incomplete. Even in the regime where the elastic approximation 
is highly accurate from the energy point of view, small inelasticity
will affect the asymmetric part of the cross section. This comes about 
because the asymmetric terms depend in a subtle way on the available phase
space of the scattering. Indeed, our full calculation presented in \S IV
reveals additional terms in the expressions of $\eps_{\rm in}$ and 
$\eps_{\rm out}$. Moreover, to obtain the the correct expression 
for the asymmetric neutrino flux, the elastic cross section is
inadequate; one must incorporate the full inelastic effect in the
Boltzmann equation (see \S IV). A similar comment can be made for 
neutrino-nucleon absorption, where one must incorporate the
recoil motion of the nucleon as well as the Landau levels for the electron
in order to obtain the correct cross section (See \S V). 

Second, and more importantly, eq.~(\ref{scatt}) is incorrect, therefore
eq.~(\ref{fluxx}) is also incorrect and there is no drift flux proportional
to $U_\nu$. The problem with eq.~(\ref{scatt}) can easily be seen by 
considering detailed balance (e.g., Ref.~\cite{Lifshitz}):
The right-hand-side of eq.~(\ref{scatt})
must vanish when neutrinos are in thermal equilibrium with the matter.
Substituting $\fnu(\bk)$ and $\fnu(\bk')$ by the equilibrium 
distribution $\fnu^{(0)}(k)$ and using eq.~(\ref{elasticcross}),
we find the RHS of eq.~(\ref{scatt}) to be $\kappa\fnu^{(0)}(1-\fnu^{(0)})
(\eps_{\rm out}-\eps_{\rm in})\bo\cdot\bbhat$. It is exactly this 
violation of detailed balance that gives rise to the drift flux term in
eq.~(\ref{fluxx}). It is also clear that any physical drift flux 
must depend on the {\it deviation} from the equilibrium distribution.
 
Equation (\ref{scatt}) is the starting point of almost all astrophysical
radiative transport studies (e.g., Ref.~\cite{Chandra}). However, it is
invalid in the presence of asymmetric scattering. A crucial (but incorrect)
assumption implicit in eq.~(\ref{scatt}) is that the cross section for
scattering into the beam (propagating along $\bo$), 
$(d\kappa/d\Omega)_{\bo'\rightarrow\bo}$, is related to that for scattering
into the beam, $(d\kappa/d\Omega')_{\bo\rightarrow\bo'}$, by merely switching
$\bo$ and $\bo'$. In reality, however, the two cross sections have 
slightly different forms such that detailed balance is satisfied in 
equilibrium (see eq.~[\ref{scatter}]).
In other words, although the first (second) term
in eq.~(\ref{scatt}) represents a good approximation to the actual
probability of scattering into (out of) the beam, the error in their 
difference is significant. It will become clear in our study 
presented the following sections that to properly take into account 
of the detailed balance constraint, one must incorporate 
inelasticity----no matter how small----into the Boltzmann equation when
deriving the asymmetric neutrino flux.

\section{\bf GENERAL FORMALISM}

In this section, we set up the general framework to study 
neutrino transport 
in magnetic fields. As \S II shows, it is important to 
properly define the relevant cross sections which enter
the transport equation. The actual calculations of the 
cross sections are given in \S IV and \S V. 

The Boltzmann equation for neutrino transport is written in the form
\beqa
\frac{ \partial \fnu(\bk)}{\partial t} + \bo \cdot \nabla \fnu(\bk)
& = & \left[{\partial f_\nu(\bk)\over\partial t}\right]_\sc
+ \left[{\partial f_\nu(\bk)\over\partial t}\right]_{\rm abs}
\label{boltzmanneqn}
\eeqa
where $\bk=k\bo$ is the neutrino momentum,
both scattering and absorption collisions terms are included 
on the right-hand-side of the equation and we have suppressed the position
and time dependence in the neutrino distribution function $\fnu$. 

\subsection{The Scattering Term}

The collision term due to neutrino-nucleon scattering can be written as
(e.g., Ref.~\cite{kt90})
\beqa
\left[{\partial f_\nu(\bk)\over\partial t}\right]_\sc &=&
\sum_{ss'}\int \frac{d^3k'}{(2\pi)^3}\,\frac{d^3p}{(2\pi)^3} 
\,\frac{d^3p'}{(2\pi)^3}\,
\left(2 \pi \right)^4 \delta^4
\left( P + K - P' - K' \right)
\left| M_{ss'}(\bo,\bo') \right|^2
\nonumber \\ && \times
\Bigl[(1-f_{\nu})(1-f_N)f_N'f_{\nu}'
- f_{\nu}f_N(1-f_N')(1-f_{\nu}') \Bigr],
\eeqa
where $f_\nu=f_\nu(\bk)$ and $f_\nu'=f_\nu(\bk')$ are the initial
and final state neutrino distribution functions, $f_N=f_N(E)$ and
$f_N'=f_N(E')$ are the initial and final state nucleon distribution functions,
$P$ ($P'$) and $K$ ($K'$) are the initial (final) state nucleon and
neutrino four vectors, respectively, and $s,s'=\pm 1$ specify the 
initial and final nucleon spins. Time-reversal symmetry has been used
to relate the matrix element for scattering in and out of the beam.
Note that in the case of neutrino-proton
scattering, we neglect the Landau levels of proton, 
and therefore the proton
momentum is a well-defined quantity. This is justified because 
many Landau levels are occupied for the conditions
in a proto-neutron star, and the change in the available 
phase space due to the Landau levels is negligible.
Nucleons are always in thermal equilibrium, and the 
nucleon distribution function is given by 
\beqa
f_N(E) & = & \frac{1}{\exp \left[ (E-\mu_N)/T \right] + 1},
\eeqa 
where $\mu_N$ is the nucleon chemical potential (excluding rest mass).
As the neutrinos exchange energy with matter only through the weak 
interactions, their distribution can deviate from equilibrium, especially in
the outer layer of the star.

The scattering rate can be rewritten in a more conventionial form as
follows. Define the differential cross section (per unit volume) to be
\beqa
\frac{ d\Gamma}{dk'd\Omega'} & = & 
\frac{ {k'}^2}{(2\pi)^3} \sum_{ss'}
\left| M_{ss'}(\bo,\bo') \right|^2 S_{ss'}(q_0,q),
\label{xsection}
\eeqa
where the ``nucleon response function", $S_{ss'}(q_0,q)$,
is
\beqa
S_{ss'}(q_0,q) & = & 
\int\!\! \frac{ d^3p}{(2\pi)^3}\frac{ d^3p'}{(2\pi)^3}
\left(2 \pi \right)^4 
\delta^4 \left( P + K - P' - K' \right) 
\fN(1-\fN').
\label{fresponse}
\eeqa
Here we have defined the energy transfer $q_0$ and the
momentum transfer $q$ via:
\be
q_0=k-k',~~~q=|\bk-\bk'|=(k^2+{k'}^2-2kk'\bo \cdot \bo')^{1/2}.
\ee
For nucleons in thermal equilibrium, energy conservation gives
\beqa
\frac{(1-\fN)\fN '}{\fN(1-\fN ')} & = & 
\exp[(E-E')/T]= \exp[-(k-k')/T]= \exp(-q_0/T).
\eeqa
Using this expression
to relate the scattering into and out of the beam and plugging in
the differential cross section, the scattering rate becomes
\beqa
&& \left[{\partial \fnu(\bk)\over\partial t}\right]_\sc =
\int_0^{\infty}\!\! dk'\int\!\! d\Omega' \frac{ d\Gamma}{dk'd\Omega'}
\left[ e^{-q_0/\kt}(1-\fnu)\fnu'- \fnu (1-\fnu')
 \right].
\label{scatteringrate}
\eeqa 
Note that one can also explicitly define a differential cross-section,
$[d\Gamma/(dk'd\Omega')]_{+}$, for scattering into the beam by 
writing
\beqa
&& \left[{\partial f(\bk)\over\partial t}\right]_\sc =
\int_0^{\infty}\!\! dk'\int\!\! d\Omega' 
\left[\left(\frac{ d\Gamma}{dk'd\Omega}\right)_+
(1-\fnu)\fnu'
-\left(\frac{ d\Gamma}{dk'd\Omega'}\right)
\fnu (1-\fnu') \right].
\label{scatter}
\eeqa 
Clearly we have
\be
\left(\frac{ d\Gamma}{dk'd\Omega}\right)_+
=e^{-q_0/T}\left(\frac{ d\Gamma}{dk'd\Omega'}\right).
\label{relation}\ee

All the microphysics is now contained in $d\Gamma/(dk' d\Omega')$. 
It depends only on $k$, $k'$, and $\bo \cdot \bo'$,
or equivalently, $k$, $q_0$ and $q$, as can be
seen from eq.~(\ref{fresponse}). 
Note that if one sets $q_0/\kt =0$, then
the neutrino Fermi blocking terms proportional to 
$f_{\nu}(\bk)f_{\nu}(\bk')$ cancel (e.g., Ref.~\cite{lp76}). Hence, the
neutrino degeneracy does not enter the scattering rate if the 
elastic limit is taken in the phase space integrals.

It is instructive to compare eq.~(\ref{scatter}) with the (wrong)
eq.~(\ref{scatt}). In general, the cross section for scattering into 
the beam has a different form as that for scattering out of the beam.
This difference, even though numerically small, 
is essential for maintaining detailed balance
in thermal equilibrium (see \S III.C below). 
Also, as we show in the next few sections,
one cannot trivially take the elastic 
limit in eq.~(\ref{relation}), because this will lead to zero drift flux
even when the neutrino distribution deviates from thermal equilibrium.

\subsection{The Absorption Term}

The collision term in eq.~(\ref{boltzmanneqn}) for absorption/emission
is
\beqa
\left[{\partial f_\nu(\bk)\over\partial t}\right]_{\rm abs} 
& = & \int\! d\Pi_e d\Pi_n d\Pi_p\, W_{if}^{\rm (abs)}
\left[f_p f_e(1-f_n)(1-\fnu)-(1-f_p)(1-f_e)\fnu f_n \right]
\label{absorb}
\eeqa
where $W_{if}^{\rm (abs)}$
is the transition rate (S-matrix squared 
divided by time) for absorption,
and we have used time-reversal invariance of the weak Hamiltonian
to relate the S-matrix for each direction. The notation $d\Pi$
denotes sum of states (including spins). 
Note that since we will include Landau levels
for electrons and protons, $d\Pi_{e,p}$ is not equal to 
$d^3p_{e,p}/(2\pi)^3$ (see Appendix C). The components of the
transverse momentum (perpendicular to the magnetic field) are
not conserved, although we
still have conservation of the $z$-momentum and energy conservation
\beqa
k + E_n + Q & = & E_e + E_p,
\eeqa
where $Q$ is the mass difference between neutron and proton
(Recall that $E_n$ and $E_p$ do not include rest-mass).  
Since the electron, proton, and neutron are in thermal equilibrium with 
Fermi-Dirac distributions, we have the equality
\beqa
\frac{f_p f_e (1-f_n)}{(1-f_p)(1-f_e)f_n} & = & 
\exp \left[ - \left( \frac{k-\mu_{\nu}}{T} \right) \right], 
\eeqa
where we have defined the neutrino chemical potential
\beqa
\mu_{\nu} & \equiv & \mu_e + \mu_p - \mu_n - Q.
\eeqa
Equation (\ref{absorb}) then takes on the standard form
\beqa
\left[{\partial f_\nu(\bk)\over\partial t}\right]_{\rm abs} 
& = & - \kabsstar \delta\fnu,
\label{absorb2}\eeqa
where $\delta f_\nu=f_\nu(\bk)-f_\nu^{(0)}(k)$ measures the deviation of
neutrino distribution from thermal equilibrium (see below). Here
\beqa
\kabsstar & = &  \kabs \left[1+\exp \left(\frac{\mu_{\nu}-k}{T}\right) 
\right] ,
\label{kappastar}\eeqa
and $\kabs$ is the absorption opacity:
\beqa
\kabs & = & \int\! d\Pi_e d\Pi_n d\Pi_p\,W_{if}^{(\rm abs)}
(1-f_p)(1-f_e) f_n.
\label{defkappaa}
\eeqa
The factor $[1+e^{(\mu_\nu-k)/T}]$ in $\kabsstar$ 
takes into account the effect of stimulated 
absorption (e.g., Ref.~\cite{Imshennik73}).

\subsection{Detailed Balance}

In thermal equilibrium, the neutrino has the Fermi-Dirac 
distribution function
\beqa
\fnu & =& \fnu^{(0)}(k) = \frac{1}{ \exp[(k-\mu_{\nu})/\kt]+1 }.
\eeqa
We then find
\beqa
\frac{\fnu^{(0)} (1-{\fnu'}^{(0)})}{(1-\fnu^{(0)}){\fnu'}^{(0)}} & = & 
\exp[ (k'-k)/T ]= \exp(-q_0/\kt),
\eeqa
so that the scattering rate in eq.~($\ref{scatteringrate}$) is zero,
as required by detailed balance. Similarly, for $f_\nu=f_\nu^{(0)}$,
the absorption rate in eq.~(\ref{absorb2}) vanishes. 
Therefore the only nonzero contribution to 
$(\partial f_\nu/\partial t)_{\rm sc}$ and 
$(\partial f_\nu/\partial t)_{\rm abs}$ must be proportional to
the deviation of neutrino distribution from thermal equilibrium.
This implies that there is no drift flux along the magnetic field
proportional to $f_\nu^{(0)}$ (see \S II.A).

As noted before (\S III.A), when one writes the scattering rate in the form
of eq.~(\ref{scatter}), it is essential to distinguish 
$(d\Gamma/dk'd\Omega)_+$ from $(d\Gamma/dk'd\Omega')$ in order to
satisfy detailed balance.

\subsection{Deviation from Thermal Equilibrium}

In order to calculate the size of the asymmetric flux, we
must consider the deviation of neutrino distribution 
from thermal equilibrium:
\be
\delta f_\nu(\bk)=f_\nu(\bk)-f_\nu^{(0)}(k).
\ee
For $\nu_e$ and $\bar\nu_e$, the neutrino-matter energy exchange is
mediated primarily by absorption and emission via the URCA processes
(and to a lesser extent by electron-neutrino scattering), while
neutrino transport is affected by both absorption and $\nu-N$ scattering. 
The result is that the decoupling sphere of electron type 
neutrinos lies only
slightly deeper than the neutrinosphere and only a small region over
which the cumulative asymmetry can develop. 
For $\mu$ and $\tau$ neutrinos, the transport opacity is primarily 
from $\nu-N$ scattering, while energy exchange is due to inelastic
$\nu-e^{-}$ scattering. As a consequence, the decoupling layer is much
deeper than the neutrinosphere and the asymmetric flux has a large
optical depth over which to develop. Unfortunately, as
we will show, the flux asymmetry due to the $\mu$ and $\tau$ neutrinos
is cancelled by that from the corresponding antineutrinos.

For the purpose of deriving the moment equations of neutrino transport,
we shall expand $\delta\fnu(\bk)$ in spherical harmonics up to quadrupole 
order as
\beqa
\delta \fnu(\bk) & \equiv &   g(k) + 3 \bo \cdot \bh(k)
+ \frac{10}{3}  I_{ij}(k) \calp_{ij}(\bo) + ....
\label{expandfnu}
\eeqa
where 
\be
\calp_{ij} ={1\over 2} (3\Omega_i \Omega_j - \delta_{ij}),
\ee
(The components of $\calp_{ij}$ can be explicitly expressed solely in terms
of the quadrupole spherical harmonics $Y_{2m}$).  
In eq.~(\ref{expandfnu}), 
$g(k)$ is the spherically symmetric deviation from equilibrium,
$\bh(k)$ represents the dipole deviation which leads to the flux, 
and $I_{ij}(k)$ is the tensor describing the pressure asymmetry.
Since $\calp_{ij}$ is symmetric we can choose $I_{ij}$ to be symmetric,
leaving six independent elements. Moreover, we shall choose
$I_{ij}$ to be traceless (The nonzero trace can always be incorporated
into $g_\nu$).

To be explicit about the physical meaning of each component of 
$\delta\fnu$, one can relate $g,~\bh$, and $~I_{ij}$ to more commonly used
quantities. The energy density per unit energy interval is 
\beqa 
U_{k} & = & \int \frac{k^2 d\Omega}{(2\pi)^3} k \fnu 
= \frac{4\pi k^3 }{(2\pi)^3} \left[ \fnu^{(0)}(k) + g(k) \right]. 
\eeqa 
The energy flux per unit energy interval is 
\beqa 
{\bf F}_{k} & = & \int \frac{k^2 d\Omega}{(2\pi)^3} k \bo \fnu  
 =  \frac{4\pi k^3 }{(2\pi)^3} \bh(k). 
\eeqa 
Using the indentity
\beqa
\int \frac{d\Omega}{4\pi} \Omega_i \Omega_j \calp_{kl} & = & 
- \frac{1}{15} \delta_{ij} \delta_{kl} + \frac{1}{10} \left(  
\delta_{ik}\delta_{lj} + \delta_{il}\delta_{jk} \right)
\eeqa
and the fact that $I_{ij}$ is traceless, 
the pressure tensor per unit energy interval is
\beqa 
\left[ {\bf P}_{k} \right]_{ij} & = & 
\int \frac{k^2 d\Omega}{(2\pi)^3} k \Omega_i \Omega_j \fnu
= \frac{4\pi k^3 }{(2\pi)^3}  \left[ \frac{1}{3} \delta_{ij}
\left( \fnu^{(0)}(k) + g(k) \right) + \frac{2}{3} I_{ij}(k) \right]
\nonumber \\ & = & 
\frac{1}{3} \delta_{ij} U_{\bk} 
+ \frac{2}{3} \frac{4\pi k^3 }{(2\pi)^3} I_{ij}(k). 
\eeqa 
Thus $I_{ij}$ is the anisotropic portion of the pressure tensor.

Note that we have truncated our expansion at the
quadrupole order angular dependence, since each successive term
will be smaller than the previous by a factor of \cite{mihalas2}
$\sim \tau^{-1}$ (where $\tau$ is the optical depth).
It will be shown that $h \propto \fnu^{(0)}\tau^{-1}$, and
$g$ and $I_{ij}$ both scale as $\fnu^{(0)} \tau^{-2}$ (for $B=0$).
The $l$-th spherical harmonic would have
coefficients which scale as $\fnu^{(0)} \tau^{-l}$.

\section{\bf NEUTRINO-NUCLEON SCATTERING}

In this section, we calculate the differential cross-section
for $\nu-N$ scattering in magnetic fields for general conditions
of nucleons (non-relativistic but arbitrary degeneracy).
We also obtain explicit expressions for the angular moments of the
scattering term of the Boltzmann equation. These moments are then
used in (\S VI) to obtain the neutrino flux, as well as the spherical and
quadrupole deviations from thermodynamic equilibrium, on which the
asymmetric flux depends.

We approximate the nucleons as nonrelativistic particles with energy 
(excluding rest mass)
\be
E(\bp,s)=  {p^2\over 2m} - s\mu_B B,
\label{energy}\ee
where $B$ is the magnetic field strength,
$s=\pm 1$ is the nucleon spin projection along the $z$-axis (in the direction
$\bbhat$),
and $\mu_B=ge\hbar/(2mc)$ is the nucleon magnetic moment 
($g_n=-1.913$ for the neutron and $g_p=2.793$ for proton).
In this section, we shall omit the label $n$ or $p$ whenever possible,
denoting final-state quantities by a prime.
Also, the quantization of the proton energy levels
and nucleon-nucleon interactions are neglected (see \S II.A),
although we shall include proton Landau levels in our calculation of
the absorption opacity (\S V). 


\subsection{The Differential Cross Section}
\label{crosssection}

The differential cross section, defined in eq.~(\ref{xsection}),
can be evaluated analytically to linear order in $B$ for 
general conditions of nucleons. 
Since $\mu_B B =3.15\times 10^{-4}gB_{14}$~MeV (where 
$B_{14}$ is the field strength in units of $10^{14}$~G) 
is much smaller than the temperature or nucleon Fermi energy
in the proto-neutron star, an expansion in the lowest nonvanishing 
power of $B$ is an excellent approximation. 

The matrix element, $|M_{ss'}(\bo,\bo')|^2$, 
for the case in which both the initial and final nucleon states are polarized
has been derived in appendix A with the result (eq.~[\ref{m2}]):
\beqa
\left| M_{ss'}(\bo,\bo') \right|^2 & = &
\frac{1}{2} G_F^2 c_V^2 \left\{
\left( 1+3 \lambda^2 \right) + \left( 1-\lambda^2 \right) \bo \cdot \bo'
\right. \nonumber \\ && \left.
+ 2\lambda(\lambda+1)(s\bo+s'\bo')\cdot \bbhat
- 2\lambda(\lambda-1)(s\bo'+s'\bo)\cdot \bbhat
\right. \nonumber \\ && \left.
+ ss' \left[ \left( 1-\lambda^2 \right)(1+\bo \cdot \bo')
+4\lambda^2\bo \cdot \bbhat \bo' \cdot \bbhat \right]\right\}
\label{m21}
\eeqa
where $G_F,~c_V, and ~\lambda=c_A/c_V$ are the weak interaction
constants defined in appendix A. 
The nucleon response function, defined in eq.~(\ref{fresponse}), 
has been calculated in Appendix B. It can be written as
$S_{ss'}=S_0+\delta S_{ss'}$, where $S_0$ is the spin-independent
$B=0$ result and $\delta S_{ss'}$ is the correction linear in $B$. 
Combining the expressions for $|M_{ss}(\bo,\bo')|^2$ and
$S_{ss'}$ into eq.~(\ref{xsection}), we find:
\beqa
\frac{ d\Gamma}{dk'd\Omega'} & = &
A_0 \left(k,k',\mu' \right) +
\delta A_{+} \left(k,k',\mu' \right) \bo \cdot \bbhat
+ \delta A_{-} \left(k,k',\mu' \right) \bo' \cdot \bbhat,
\label{expandedxsection}
\eeqa
where $\mu'=\bo \cdot \bo'$ (not to be confused with the nucleon 
magnetic moment, $\mu_B$, or the nucleon chemical potential
$\mu_N$). The first term in eq.~(\ref{expandedxsection})
is the $B=0$ result:
\beqa
A_0 \left(k,k',\mu'\right) & = &
\frac{ {k'}^2}{(2\pi)^3} \sum_{s,s'}
\left| M_{ss'}(\bo,\bo') \right|^2 S_0(q_0,q)
\nonumber \\ & = &
\frac{ {k'}^2}{(2\pi)^3}
2 G_F^2 \cv^2
\left[ \left( 1 + 3 \lambda^2\right)
+ \left( 1  - \lambda^2  \right)\mu' \right] S_0(q_0,q),
\label{A0}
\eeqa
with
\beqa
S_0(q_0,q) & = & \frac{m^2 \kt}{2\pi q}
\frac{1}{1-e^{-z}}
\ln \left[ \frac{1+\exp(-x_0)}{1+\exp(-x_0-z)} \right],
\eeqa
and we have defined
\beqa
x_0 & = &  \frac{(q_0-q^2/2m)^2}{4\kt(q^2/2m)} - \frac{\mu_N}{\kt}
\mbox{\ \ \ \ \ {\rm and} \ \ \ \ \ } z=\frac{q_0}{T}.
\eeqa 
The second and third terms in eq.~(\ref{expandedxsection})
correspond to the corrections arising from nonzero $B$:
\beqa
\delta A_{+} \bo \cdot \bbhat
+ \delta A_{-}  \bo' \cdot \bbhat
&=& \frac{ {k'}^2}{(2\pi)^3} \sum_{s,s'}
\left| M_{ss'}(\bo,\bo') \right|^2 \delta S_{ss'}(q_0,q),
\eeqa
with the coefficients:
\beqa
\delta A_{\pm} \left(k,k',\mu' \right) & = &
 \frac{ {k'}^2}{(2\pi)^3} \frac{2G_F^2\cv^2m^2\lambda \mu_B B}{\pi q}
\nonumber \\ && \times
\frac{ 1}
{ \left[ \exp(x_0)+1 \right] \left[ 1 + \exp(-x_0-z) \right] } 
\left( 1 \pm \lambda \frac{2mq_0}{q^2} \right).
\label{Apm}
\eeqa

The reason for writing the cross section in the form of
eq.~(\ref{expandedxsection}) is that the angular dependence
needed to find the moment equations (see \S IV.C) is now manifest.
Note that the cross section in eq.~(\ref{expandedxsection}) exhibits parity
violation. If the parity operation is taken, the vectors $\bo$ and
$\bo'$ reverse sign and the pseudovector $\bbhat$ keeps the same sign
so that the cross section does not retain the same form. 
Also note that the cross section for scattering from the state
$\bo$ to the state $\bo'$ does not have the same numerical value as
the reverse process. However, this does not mean that time reversal
invariance is violated. The inequality arises from averaging the 
matrix element over the nucleon distribution functions.
Indeed, the matrix element in eq.~(\ref{m21}) can be explicitly
shown to satisfy time reversal invariance by simultaneously interchanging
all initial and final state labels.

\subsection{The Differential Cross Section: Nondegenerate Nucleon Limit}

Even after expanding the cross section in Eqs. (\ref{A0}) and 
(\ref{Apm}) for small magnetic fields, the expressions are still 
quite difficult to evaluate in general. However, as 
discussed in \S III, asymmetric drift flux can develop only when 
the neutrino distribution deviates from thermal equilibrium (i.e.,
above the decoupling sphere). This occurs in the regime where
nucleons are nondegenerate (at density
$\rho \sim 10^{12}-10^{13}$~g~cm$^{-3}$).
In this subsection we derive
simplified expressions of $A_0$ and $\delta A_\pm$ which will be useful
for obtaining the angular moments of the scattering term (\S IV.C and D)
and neutrino flux. 
 
For degenerate nucleons, the characteristic neutrino energy transfer
in each scattering is of order $q_0\sim k(T/m)^{1/2}\ll k$. The cross
section peaks sharply around $k'=k$, and we can evaluate 
$A_0,~\delta A_\pm$ in a series in the small parameter $(T/m)^{1/2}$.
Define the dimensionless quantities
\beqa
\epsilon=[4(1-\mu')T/m]^{1/2},~~~~~
u={k'-k\over \epsilon k},
\label{defineeps}\eeqa
so that the range of $u$, the dimensionless neutrino energy,
is from $-1/\epsilon$ to $\infty$. 
Using the expansion of the nucleon response function derived in Appendix B,
we have, to linear order in $\eps$,
\beqa
A_0 & \simeq & \left( \frac{G_F c_V k}{2\pi} \right)^2
\left[ 1 + 3\lambda^2 + (1-\lambda^2)\mu' \right]
\frac{n}{k\epsilon \pi^{1/2}} e^{-u^2}
\left[ 1 + \frac{3}{2} \epsilon u + \epsilon u^3 - 
2 \frac{ k(1-\mu')}{\epsilon m} u\right],
\label{elasticA0}
\eeqa
and
\beqa
\delta A_{\pm} 
& \simeq & 
\left( \frac{G_F c_V k}{2\pi} \right)^2
\frac{2\mu_B B}{T} \lambda \frac{n}{k\epsilon \pi^{1/2}}
e^{-u^2} 
\left[ 1 + \frac{3}{2} \epsilon u + \epsilon u^3 -
2 \frac{ k(1-\mu')}{\epsilon m} u
\right. \nonumber \\ & &\mp  \left.
\frac{\lambda \epsilon m} {k(1-\mu')}
\left( u + \frac{1}{2} \epsilon u^2 + \epsilon u^4 -
2 \frac{ k(1-\mu')}{\epsilon m} u^2  \right)\right]. 
\label{elasticdeltaApm}
\eeqa
In deriving these expressions we have used the the $B=0$ equation
\be
\exp\left({\mu_N\over T}\right)=n\left({2\pi^3\over m^3T^3}\right)^{1/2}
\label{chemical}\ee
to relate the nucleon chemical potential $\mu_N$ to its
number density $n$ (The corrections due to finite $B$ are of order $B^2$).
These expansions of $A_0,~\delta A_\pm$
are valid under the conditions (see Appendix B for details)
$T\ll m$,
$k \ll (mT)^{1/2}$,
$\mu_B B \ll T$ and
$k\simgreat k_{min}= \mu_B B (m/T)^{1/2}
\simeq 10^{-2}|g| B_{14} T^{-1/2}$~MeV.
These conditions are satisfied for the conditions 
of interest in our study\footnote{ For $k\simless k_{min}$, the
expansions leading to eqs. (\ref{elasticA0}) and (\ref{elasticdeltaApm})
are no longer valid, so that different approximations must be made 
(the $k \rightarrow 0$ limit).
Since this is a relatively small range of neutrino energy, we ignore
this complication here.}.

Note that dimensionally,
$\delta A_\pm$ is smaller than $A_0$ by a factor
of order $\mu_BB/T$, but also note that 
the quantity $\epsilon m/k$ (which appears on the second line of
eq.~[\ref{elasticdeltaApm}]) is of order $\sqrt{mT}/k \sim \sqrt{m/T}$, 
which can be quite large. 
This point will be important when we consider the size of the neutrino 
drift flux.

\subsection{\bf Moments of the Scattering Rate}
\label{diffusionmoments}

To derive the expression for the neutrino flux, one needs to 
take the angular moments of the Boltzmann transport equation
(i.e., multiply the equation by some power of $\bo$ and then integrate
over $d\Omega$). In this subsection we derive the general expressions for 
the moments of the scattering rate in the Boltzmann equation.
In the next subsection we shall evaluate these expressions explicitly for the
regime when the nucleons are nondegenerate and the scattering is 
approximately elastic.


We first write the scattering rate (eq.~[\ref{scatteringrate}])
in terms of $\delta f_\nu\equiv f_\nu-f_\nu^{(0)}$:
\beqa
&& \left[{\partial \fnu(\bk)\over\partial t}\right]_\sc =
\int_0^{\infty}\!\! dk'\int\!\! d\Omega' \frac{ d\Gamma}{dk'd\Omega'}
\left[ C(k,k')\delta\fnu' + D(k,k')\delta\fnu + 
E(k,k')\delta\fnu'\delta\fnu\right],
\label{expandedrate}
\eeqa
where the $C,~D,~E$ coefficients, and their $q_0/T \ll 1$ expansions,
are
\beqa
C(k,k') & = & e^{-q_0/\kt} \left(1-\fnu^{(0)} \right) + \fnu^{(0)}
\simeq 1 + \left( - \frac{q_0}{T} + \frac{q_0^2}{2T^2} \right)
(1-\fnu^{(0)}),\label{CandC}\\
D(k,k') & = & - \left[  e^{-q_0/\kt}{\fnu^{(0)}}' +  1-{\fnu^{(0)}}' \right]
\simeq -1 + \frac{q_0}{T}\fnu^{(0)} - 
\frac{q_0^2}{2T^2} \left( \fnu^{(0)}
+ 2T \frac{\partial \fnu^{(0)}}{\partial k} \right),
\label{CandD}\\
E(k,k') & = & 1 - e^{-q_0/\kt} \simeq \frac{q_0}{T} - \frac{q_0^2}{2T^2}.
\eeqa
The nonlinear terms $E(k,k')\delta\fnu\delta\fnu'$ in 
eq.~(\ref{expandedrate}) will be dropped since we consider the regime where
the deviation of $f_\nu$ from thermal equilibrium is relatively small
(the regime where the moment formalism is valid). 


Plugging eqs.~($\ref{expandfnu}$) and ($\ref{expandedxsection}$) into
eq.~($\ref{expandedrate}$) and then performing the azimuthal integrals
using the indentities
\beqa
&& \int_0^{2\pi} \frac{d\phi'}{2\pi} \Omega'_i  =  \mu' \Omega_i
\nonumber \\
&& \int_0^{2\pi} \frac{d\phi'}{2\pi} \Omega'_i \Omega'_j  = 
P_2(\mu') \Omega_i \Omega_j + \frac{1}{3} \left(1-P_2(\mu') \right) 
\delta_{ij},
\nonumber \\ &&
\int_0^{2\pi} \frac{d\phi'}{2\pi} \calp_{ij}(\bo')  =  
P_2(\mu') \calp_{ij}(\bo)
\nonumber \\ &&
\int_0^{2\pi} \frac{d\phi'}{2\pi} \Omega'_i \calp_{jk}(\bo')  =  
\frac{3}{5} \calp_{ijk}(\bo) - \frac{1}{5} \mu' \Omega_i \delta_{jk}
+ \frac{3}{10} \mu'
\left[ \Omega_j \delta_{ki} + \Omega_k \delta_{ij} \right],
\eeqa
where the three-index tensor equivalent to the $Y_{3m}$ is defined
by $\calp_{ijk}(\bo) = (5 \Omega_i \Omega_j \Omega_k - 
\Omega_i \delta_{jk} - 
\Omega_j \delta_{ki}- \Omega_k \delta_{ij})/2$, we find
\beqa
\left[{\partial \fnu(\bk)\over\partial t}\right]_\sc =&&
\int_0^{\infty} dk' \int_{-1}^{1} d\mu' \int_{0}^{2\pi} d\phi'
\left[ A_0 + \delta A_{+} \bo \cdot \bbhat + \delta A_{-} 
\bo' \cdot \bbhat \right]\nonumber \\ 
&&\times
\left[ C \left( g' + 3 \bo' \cdot \bh'
+ \frac{10}{3}  I'_{ij} \calp'_{ij} \right) 
+ D \left( g + 3 \bo \cdot \bh
+ \frac{10}{3}  I_{ij} \calp_{ij} \right)
\right]\nonumber \\ 
=&&2\pi \int_0^{\infty} dk' \int_{-1}^{1} d\mu'
\left[ A_0 + \delta A_{+} \bo \cdot \bbhat \right]
\Biggl\{ \left(Cg'+Dg \right) 
+ 3 \left( C\mu'\bh' + D\bh \right) \cdot \bo \nonumber \\ 
& &+ \frac{10}{3} \left[ CP_2(\mu')I'_{ij} + 
DI_{ij} \right] \calp_{ij} \Biggr\}\nonumber \\ 
& &+ 2\pi \int_0^{\infty} dk' \int_{-1}^{1} d\mu' \delta A_{-}
\Biggl\{ \mu' \left(Cg'+Dg \right) \bo \cdot \bbhat
+ 3C P_2(\mu') (\bo \cdot \bbhat) (\bo \cdot \bh') \nonumber\\
&&+ C \left[1-P_2(\mu')\right]\bbhat \cdot \bh' 
+ 3D\mu' (\bo \cdot \bbhat)(\bo \cdot \bh) \nonumber\\
&&+ \frac{10}{3} C \bhat_iI'_{jk} \left[\frac{3}{5} \calp_{ijk}
- \frac{1}{5} \mu'\Omega_i \delta_{jk} 
+ \frac{3}{10}\mu' \left( \Omega_j \delta_{ki} + \Omega_k \delta_{ij}
 \right) \right]\nonumber\\
&&+ \frac{10}{3}D\mu' I_{ij} \calp_{ij} \bo \cdot \bbhat \Biggr\}.
\label{bigmess}
\eeqa

To calculate the moments of the scattering rate, the following identities
are needed:
\beqa
&& \int \frac{d\Omega}{4\pi} \Omega_i \Omega_j =  \frac{1}{3} \delta_{ij}
\nonumber \\ &&
\int \frac{d\Omega}{4\pi} \calp_{ij}  =  0
\nonumber \\ && 
\int \frac{d\Omega}{4\pi}  \calp_{ij}  \calp_{kl} =  
- \frac{1}{10} \delta_{ij}\delta_{kl} 
+ \frac{3}{20} \left(\delta_{ik}\delta_{lj} + \delta_{il}\delta_{jk}
 \right)
\nonumber \\ &&
\int \frac{d\Omega}{4\pi} \Omega_i \Omega_j \calp_{kl}  =  
- \frac{1}{15} \delta_{ij}\delta_{kl} 
+ \frac{1}{10} \left(\delta_{ik}\delta_{lj} + \delta_{il}\delta_{jk}
 \right)
\nonumber \\ &&
\int \frac{d\Omega}{4\pi} \calp_{ijk} = 0
\nonumber \\ && 
\int \frac{d\Omega}{4\pi} \calp_{ijk} \Omega_l = 0.
\label{angularintegrals}
\eeqa
Also note that any integral of an odd number of $\Omega_i$'s
over the solid angle gives zero. 
The zeroth moment of eq.~(\ref{bigmess}) is
\beqa
\int \frac{d\Omega}{4\pi} 
\left[{\partial \fnu(\bk)\over\partial t}\right]_\sc
& = & 2\pi \int_0^{\infty} dk' \int_{-1}^{1} d\mu' \Bigl[
A_0 \left( Cg' + Dg \right) 
\nonumber \\ & + & 
\delta A_{+} (C\mu'\bh' + D\bh) \cdot \bbhat + 
\delta A_{-} (D\mu'\bh + C\bh') \cdot \bbhat 
\Bigr],
\label{mom0}\eeqa
the first moment is
\beqa
&& \int \frac{d\Omega}{4\pi} \Omega_i 
\left[{\partial \fnu(\bk)\over\partial t}\right]_\sc  =
2\pi \int_0^{\infty} dk' \int_{-1}^{1} d\mu'
A_0 \left[ C\mu' h'_i + Dh_i \right]
\nonumber \\ &&~~~~~~+
 \frac{2\pi}{3} \int_0^{\infty} dk' \int_{-1}^{1} d\mu'
 \left( Cg' + Dg \right)\left( \delta A_{+} + 
\mu' \delta A_{-} \right)  \bhat_i
\nonumber \\ &&~~~~~~+ 
\frac{4\pi}{3} \int_0^{\infty} dk' \int_{-1}^{1} d\mu'
\left[ \delta A_{+} \left( CP_2(\mu')I'_{ij}+DI_{ij}\right)
+ \mu'\delta A_{-} \left( CI'_{ij}+DI_{ij} \right) \right] \bhat_j,
\label{mom1}\eeqa
and the second moment is
\beqa
&& \int \frac{d\Omega}{4\pi} \calp_{ij}(\bo)
\left[{\partial \fnu(\bk)\over\partial t}\right]_\sc =
2\pi \int_0^{\infty} dk' \int_{-1}^{1} d\mu'
A_0 \left[ C P_2(\mu') I'_{ij} + D I_{ij}  \right]
\nonumber \\ && ~~~~~+
2\pi \frac{3}{10} \int_0^{\infty} dk' \int_{-1}^{1} d\mu'
\left[C \left( \mu'\delta A_{+} + P_2(\mu') \delta A_{-}  \right) 
\left( h'_i\bhat_j + h'_j\bhat_i - 
\frac{2}{3} \delta_{ij} \bh' \cdot \bbhat \right)
\right. \nonumber \\ && ~~~~~+ \left.
D \left( \delta A_{+} +  \mu' \delta A_{-} \right)
\left( h_i\bhat_j + h_j\bhat_i - 
\frac{2}{3} \delta_{ij} \bh \cdot \bbhat \right) \right].
\label{mom2}\eeqa

\subsection{Moments of the Scattering Rate: Nondegenerate Nucleon Limit}


We now evaluate the moments in eqs.~(\ref{mom0})-(\ref{mom2})
explicitly for nondegenerate nucleons (near the stellar surface)
using the expressions of $A_0$ and $\delta A_\pm$ 
(eqs.~[\ref{elasticA0}] and [\ref{elasticdeltaApm}]) as derived in
\S IV.B for small inelasticity. 
After substituting eqs.~(\ref{CandC}) and (\ref{CandD}), it will 
be necessary to evaluate moments of $q_0/T$ against $A_0$ and 
$\delta A_{\pm}$. These moments are defined as
\beqa
&& M_0^n  \equiv  \int_{0}^{\infty} dk' A_0(k,k',\mu')
\left(\frac{q_0}{T} \right)^n
= \epsilon k \left(\frac{-\epsilon k}{T} \right)^n
 \int_{-\infty}^{\infty} du A_0 u^n
\eeqa
and
\beqa
&& \delta M_{\pm}^n \equiv  \int_{0}^{\infty} dk'
\delta A_{\pm}
\left(\frac{q_0}{T} \right)^n
= \epsilon k \left(\frac{-\epsilon k}{T} \right)^n
 \int_{-\infty}^{\infty} du \delta A_{\pm} u^n,
\eeqa
where we have used the eq.~(\ref{defineeps}) and
in the $du$ integrals we have extended the lower limit $(-\eps^{-1})$
to $-\infty$ (since $\eps\ll 1$).
Only the following values will be needed
\footnote{
In $A_0$ (eq.~[\ref{elasticA0}]), the leading order term in $\epsilon$ is an
even function of $u$, and the higher order term is an odd function of $u$. 
Hence, when $q_0^n\propto u^n$ is integrated against $A_0$ to find the
moments $M_0^n$, the $n=0$ moment is larger than the $n=1,2$ moments by a
factor of $\sim k/m$, which are larger than the $n=3,4$ moments by a factor
of $\sim k/m$, etc. On the other hand, the leading order term in
$\delta A_{\pm}$ is an odd function of $u$. The moments of $q_0^n$
against $\delta A_{\pm}$, called $\delta M^n_{\pm}$, will then have 
the $n=0,1$ terms larger than $n=2,3$ by a factor of $k/m$ and so on.}: 
\beqa
 M_0^0 & = & \left( \frac{G_F c_V k}{2\pi} \right)^2 n
\left[ 1 + 3\lambda^2 + (1-\lambda^2)\mu' \right]
\nonumber \\
 M_0^1 & = &  \left( \frac{G_F c_V k}{2\pi} \right)^2 n
\left[ 1 + 3\lambda^2 + (1-\lambda^2)\mu' \right]
\left[ \frac{k}{m}(1-\mu') \left(\frac{k}{T}-6 \right) \right]
\nonumber \\
M_0^2 &= &  \left( \frac{G_F c_V k}{2\pi} \right)^2 n
\left[ 1 + 3\lambda^2 + (1-\lambda^2)\mu' \right]
\left[ 2 \frac{k^2}{mT}(1-\mu') \right]
\nonumber \\
\delta M_{\pm}^0 & = &
\left( \frac{G_F c_V k}{2\pi} \right)^2 n  \frac{2\mu_B B}{T}
\left(\lambda \pm \lambda^2 \mp 4\lambda^2 \frac{T}{k} \right)
\nonumber \\
\delta M_{\pm}^1  & = &
\pm \left( \frac{G_F c_V k}{2\pi} \right)^2 n  \frac{2\mu_B B}{T}
 \ 2\lambda^2.
\label{inelasticitymoments}
\eeqa

Let the notation ${\cal O}(n)$ mean a term which contains a factor of
$(q_0/T)^n$ in the integrands of eqs.~(\ref{mom0})-(\ref{mom2}).
In the zeroth moment, the ${\cal O}(0)$ term is zero and 
$M_0^1$ is of the same size as $M_0^2$ so we need to expand to 
${\cal O}(2)$ for the $B=0$ piece. For the $B \neq 0$ piece, 
the ${\cal O}(0)$ term is not zero, but $\delta M_{\pm}^0$ is 
of the same size as $\delta M_{\pm}^1$ so we need to expand to 
${\cal O}(1)$. The zeroth moment is then given by
\beqa
\int \frac{d\Omega}{4\pi}
\left[{\partial \fnu(\bk)\over\partial t}\right]_\sc
 = && \kscat_0 \frac{k}{m} \Biggl\{
6 \left[ T\frac{\partial g}{\partial k} + (1-2\fnu^{(0)}) g \right]
\nonumber \\ && + 
\frac{k}{T}\left[ T^2 \frac{\partial^2 g}{\partial k^2}
+ T\frac{\partial g}{\partial k}(1-2\fnu^{(0)})
- 2 g T\frac{\partial \fnu^{(0)}}{\partial k} \right] 
\Biggr\}\nonumber \\ && + 
\epsilon_{\rm sc} \kscat_0 \left( 
T \frac{\partial \bh}{\partial k} + 4 \frac{T}{k} \bh \right)
\cdot \bbhat,
\label{scatzeromom}
\eeqa
where we have defined the zero-field scattering opacity (per unit
volume)
\beqa
\kscat_0 & = & \frac{8\pi}{3} 
\left( \frac{G_Fc_Vk}{2\pi} \right)^2
(1+5\lambda^2)n,
\eeqa
and the dimensionless asymmetry parameter
\beqa
\epsilon_{\rm sc} & = & \frac{6\lambda^2}{(1+5\lambda^2)} \frac{\mu_B B}{T}.
\label{epssc}\eeqa
For the first moment, the $B=0$ terms are nonvanishing at 
${\cal O}(0)$ so that only the $q_0/T=0$ terms are needed. The
terms involving $g$ cancel at ${\cal O}(0)$ and so only the
${\cal O}(1)$ terms are needed 
(note that $\delta M_{\pm}^1 \gg \delta M_{\pm}^2$). Lastly, both
the ${\cal O}(0)$ and the ${\cal O}(1)$ terms are needed
for the $I_{ij}$ terms since $\delta M_{\pm}^0$ is the same size as
$\delta M_{\pm}^1$. The first moment is then given by
\beqa
 \int \frac{d\Omega}{4\pi} \Omega_i
\left[{\partial \fnu(\bk)\over\partial t}\right]_\sc  = && 
- \kscat_0 h_i-
 \frac{1}{3} \epsilon_{\rm sc}\kscat_0 \Biggl\{\left[
T \frac{\partial g}{\partial k} + ( 1-2\fnu^{(0)})g \right]
\bhat_i \nonumber\\
&&+ \left(1 - 2\fnu^{(0)} +\frac{1}{\lambda} - 4 \frac{T}{k}\right)
I_{ij} \bhat_j\Biggr\}.
\label{scatfirstmom} 
\eeqa
For the second moment, the $B=0$ piece is nonzero at ${\cal O}(0)$
so that only lowest order is needed. The $B \neq 0$ terms must be
kept at both ${\cal O}(0)$ and ${\cal O}(1)$. The result is
\beqa
&& \int \frac{d\Omega}{4\pi} \calp_{ij}(\bo)
\left[{\partial \fnu(\bk)\over\partial t}\right]_\sc =
- \frac{3}{2} \left( \frac{1+3\lambda^2}{1+5\lambda^2} \right) 
\kscat_0 I_{ij}
\nonumber \\ && ~~~~~~~
- \frac{3}{20} \epsilon_{\rm sc} \kscat_0
\left(1 - 2\fnu^{(0)}+
 \frac{1}{\lambda} - 4 \frac{T}{k} \right)
\left( h_i\bhat_j + h_j\bhat_i -
\frac{2}{3} \delta_{ij} \bh \cdot \bbhat  \right).
\label{scatsecondmom}  
\eeqa


\subsection{Elastic Cross Section}
\label{comparison}



As discussed before (see \S III.A), it is essential to
retain the inelasticity in the differential cross section
in order to derive the correct neutrino flux (as we have done 
in the previous subsections). Nevertheless, from the energetics
point of view, $\nu-N$ scattering is highly elastic near the 
surface of the proto-neutron star
\footnote{In fact, the scattering is elastic (i.e., $|q_0|\ll k$)
to a good approximation in most regions of the proto-neutron star. 
The only exception is during the first second or so after core collapse, when
$\nu_e$'s are highly degenerate in the stellar core.}, 
and it is instructive consider the ``elastic'' scattering rate,
$d\Gamma/d\Omega'$, obtained by
integrating $(d\Gamma/dk'd\Omega')$ over all final neutrino energies 
$k'$. Using eqs.~(\ref{elasticA0}) and (\ref{elasticdeltaApm}), together with
(\ref{inelasticitymoments}), we find:
\beqa
\frac{ d\Gamma}{d\Omega'}   = &&   \int_0^{\infty} dk' 
\frac{ d\Gamma}{dk'd\Omega'} = M_0^0 + \delta M_{+}^0 \bo \cdot \bbhat
+ \delta M_{-}^0 \bo' \cdot \bbhat
\nonumber \\  = && 
\left( \frac{G_F c_V k}{2\pi} \right)^2 n
\Biggl\{ 1 + 3\lambda^2 + (1-\lambda^2)\mu'
\nonumber \\ && +
 \frac{2\mu_B B}{T} \left[ 
\lambda \left(1+\lambda - 4\lambda \frac{T}{k} \right) \bo \cdot \bbhat
+ \lambda \left( 1-\lambda + 4\lambda \frac{T}{k} \right) \bo' \cdot \bbhat
\right] \Biggr\}.
\label{integratedrate}
\eeqa
The resulting cross-section per particle, $(d\Gamma/d\Omega)/n$,
is similar to eq.~(\ref{scatt2}) (recall that polarization $P=\mu_BB/T$
for nondegenerate nucleons), obtained by assuming complete 
elasticity of the scattering process. The difference 
involves the terms with $4\lambda T/k$ in eq.~(\ref{integratedrate}). 
These terms appear in the phase space integral of the asymmetric part of
the cross section, which can be affected by even a small inelasticity. 
In fact, for low energy neutrinos (with $k \simless 4T$),
the $T/k$ terms dominates the asymmetry in the cross section.
\footnote{Note, however, that eq.~(\ref{integratedrate}) is valid only 
for $k\go k_{min}=\mu_BB(m/T)^{1/2}$ (see the end of \S IV.B). This 
means that the maximum asymmetry is $\sim \mu_BB/k_{min}=(T/m)^{1/2}$.}



\subsection{The Anti-Neutrino Cross Section}

The expressions derived in previous sections apply only for
neutrinos. For $\bar\nu+N\rightarrow\bar\nu+N$, 
the differential cross section of the form eq.~(\ref{expandedxsection}) still
applies, except that one needs to switch the coefficients in front of
$\bo\cdot\bbhat$ and $\bo'\cdot\bbhat$. This is due to 
the crossing symmetry of the tree-level Feynmann diagram.


\section{Neutrino Absorption by Nucleons }
\label{absorption}

In this section we derive an explicit expression
the cross-section for neutrino absorption ($\nu_e+n\rightarrow p+e^-$) in
magnetic fields as formulated in \S III.B. 

In the regime where the neutrino energy is much smaller than the nucleon
rest mass, one might be tempted to consider an ``elastic'' cross-section,
obtained by treating the nucleons infinitely massive (i.e., neglecting 
nucleon recoil)\footnote{Previous authors have all neglected the recoil
effect. Refs.~\cite{Matese69,Fassio-Canuto69} focus on the rate of neutron
decay (including the effect of electron Landau levels) and do not address
the angular distribution. Ref.~\cite{Dorofeev85} discusses 
the electron contribution to the asymmetric emission in $e^-+p\rightarrow
n+\nu_e$, but the authors did not give an explicit expression.
Moreover, the cancellation of asymmetric emission and absorption 
was not considered (see \S III.B).}.
Further neglecting the effects of 
electron Landau levels and Fermi blocking, 
we obtain (e.g., Ref.~\cite{Lai98a})
\be
\sigma={1\over \pi}(G_Fc_Vk)^2\left[
(1+3\lambda^2)+2P_n\lambda(\lambda+1)\bo\cdot\bbhat\right],
\label{kappawrong}\ee
where $P_n$ is the neutron polarization,
$\bk=k\bo$ is the incident neutrino momentum, and
$G_F,~c_V,~\lambda$ are (charge-current) weak interaction 
constants (see Appendix D). This simplified treatment, however,
leads to an incomplete expression for the asymmetric part (proportional
to $\bo\cdot\bbhat$) of the cross-section\footnote{In traditional
laboratory experiments of parity violation (e.g., 
Ref.~\cite{Wu66}), the approximation 
(of neglecting nucleon recoil and electron
Landau levels) is valid because the temperature 
$T$ is much smaller than the
neutrino energy $k$. In the supernova context, $k$ is comparable to $T$
and this simple approximation leads to significant error.}.
As we show in this section, by
incorporating the effects of electron Landau levels and small inelasticity,
additional asymmetric terms which relate to electron and proton 
polarizations are revealed. These additional terms are 
important in determining asymmetric flux from the proto-neutron star. 

Note that the quantization of electron energy levels can
induce oscillatory features in the total absorption
cross-section as a function of the neutrino energy\cite{Lai98b}. 
This effect results purely from the modification of the electron phase
space due to the Landau levels (similar to the magnetization
of an electron gas at low temperatures). This oscillatory feature is
particularly prominent in the low density regime where only a few electron
Landau levels are filled. Our focus in this paper is the asymmetric part of the
cross-section, which arises from parity violation. We shall therefore restrict
to the regime where more than a few electron Landau levels are filled.
In this regime, we can replace the sum over Landau levels by an integral,
and obtain an explicit expression for the asymmetric parameter in the 
cross-section. 

In the calculation presented below, we also include the effect of 
proton Landau levels. As expected, this introduces no new term 
in the cross-section since many proton Landau levels are filled
for the typical conditions of proto-neutron stars. Our result
therefore also serves as an explicit demonstration on the validity of
using proton plane waves in calculating neutrino-nucleon opacities 
(absorption and scattering; see \S VI). It turns out that, after using 
some standard identities involving Landau wavefunctions, the calculation 
with electron and proton Landau levels is not more difficult 
than the calculation with only electron Landau levels.



\subsection{The Expression for Absorption Opacity}

The absorption opacity is defined in eq.~(\ref{defkappaa}). The energy of
a relativistic electron in a magnetic field is given by
\be
E_e=(m_e^2+2eBN_e+p_{e,z}^2)^{1/2},
\ee
where $N_e=0,~1,~2,\cdots$ is the Landau level index and
$p_{e,z}$ is the electron $z$-momentum. The other quantum numbers
specifying the electron states are: 
$\sigma_e=\pm 1$, the spin projection 
along ${\bf \Pi}=\bp+e{\bf A}$, 
and $R_e=0,~1,~2,~\cdots$, which determines the
radius of the electron guiding center (see Appendix C). Note that 
for the ground Landau level, the electron spin is opposite to the magnetic 
field, thus only the spin projection $\sigma_{e0}=-{\rm sign}(p_{e,z})$ is 
allowed. The sum over electron states is then
\beqa
\int\!d\Pi_e & = & \sum_{N_e=0}^{\infty} \sum_{\sigma_e=\pm 1} c(N_e,\sigma_e)
\sum_{R_e=0}^{R_{max}}
\int_{-\infty}^{\infty} \frac{L dp_{e,z}}{2\pi}
\eeqa
where $c(N_e,\sigma_e)=1-\delta_{N_e,0}\delta_{\sigma_e,-\sigma_{e0}}$
is zero if both $N_e=0$ and $\sigma_e=-\sigma_{e,0}$ and
one otherwise. The cutoff $R_{max} \simeq eBA/2\pi$
(the degeneracy of the Landau level)
limits the guiding center to lie within the normalization volume $V=AL$ (where
$A$ is the area). The proton energy is given by
\be
E_p={eB\over m}\left(N_p+{1\over 2}\right)-s_p\mu_{Bp}B+{p_{p,z}^2\over 2m},
\ee
where $N_p$ specifies the Landau level, $s_p=\pm 1$ is the spin 
projection along the
$z$-axis, $\mu_{Bp}$ is the proton magnetic moment, and
$p_{p,z}$ is the proton $z$-momentum. 
The summation over states for the proton takes the form
\beqa
\int\!d\Pi_p & = & \sum_{N_p=0}^{\infty} \sum_{R_p=0}^{R_{max}}
\sum_{s_p=\pm 1} \int_{-\infty}^{\infty} \frac{Ldp_{p,z}}{2\pi}
\eeqa
where $R_p$ is the quantum number for the proton guiding 
center, and $R_{max}$ is the same as for the electron.
Finally, the neutron phase space is simply that of a free particle,
with energy $E_n={\bp_n^2/(2m)}-s_n\mu_{Bn}B$.

In Appendix D, we find that the transition rate for
absorption (S-matrix squared divided by time) takes on the form
\beqa
W_{if}^{\rm (abs)} 
& = & L^{-1} V^{-2} (2\pi)^2
\delta(E_e+E_p-k-E_n-Q)\delta(p_{e,z}+p_{p,z}-k_{z}-p_{n,z})
|M|^2,
\eeqa
where the matrix element, summed over the guiding center quantum numbers
$R_e$ and $R_p$ for the electron and proton
as well as electron spin $\sigma_e$, can be written as
\beqa
\sum_{R_e=0}^{R_{max}}\sum_{R_p=0}^{R_{max}} \sum_{\sigma_e=\pm 1}
c(N_e,\sigma_e)
|M|^2 & = & \frac{G_F^2}{2} A \frac{eB}{2\pi} L_{\mu\nu} N^{\mu\nu}.
\eeqa
Here $L_{\mu\nu}$ is the lepton tensor and $N^{\mu\nu}$ is the nucleon
tensor, which takes on precisely the same form as in the zero field
case. Plugging this back into eq.~(\ref{defkappaa}) gives
\beqa
&&\kabs  =  \frac{G_F^2}{2} \frac{eB}{2\pi} 
\sum_{N_e=0}^{\infty}
\int_{-\infty}^{\infty}  \frac{ dp_{e,z}}{2\pi} (1-f_e)
\sum_{N_p=0}^{\infty} \int \frac{d^2p_{n,\perp}}{(2\pi)^2}
\sum_{s_n,s_p=\pm 1} S_{s_n s_p}
L_{\mu \nu} N^{\mu \nu},
\label{nicekappaa}
\eeqa
where we have defined a ``response function" for absorption
\beqa
S_{s_ns_p} & = & 
\int_{-\infty}^{\infty} \frac{dp_{n,z}}{2\pi} 
\int_{-\infty}^{\infty} \frac{dp_{p,z}}{2\pi}  
\nonumber \\ && \times
(2\pi)^2 \delta(E_e+E_p-k-E_n-Q) 
\delta(p_{e,z}+p_{p,z}-k_{z}-p_{n,z})
f_n(1-f_p).
\label{absresponse}
\eeqa
By integrating over the delta functions (see Appendix E), we derive
a general expression for the nucleon response function:
\beqa 
S_{s_ns_p} & = & \frac{m}{|q_z|} \left( \frac{1}{e^y+1} \right) 
\left( \frac{1}{1+e^{-y-z}} \right), 
\label{response}
\eeqa
where $q_z=k_z-p_{e,z}$, $q_{\perp}^2=eB(2N_p+1)-p_{n,\perp}^2$,
$q^2=q_z^2+q_{\perp}^2$, $q_0=k-E_e$,
\beqa
y & = & \frac{E_n-\mu_n}{T} = - \frac{\mu_n}{T}
+ \frac{ p_{n,\perp}^2}{2mT} +
\frac{ \left[q_0+Q-q^2/2m +
( \mu_{Bp}s_p-\mu_{Bn}s_n)B \right]^2}{4T(q_z^2/2m)}
- \frac{\mu_{Bn}s_nB}{T}
\eeqa
and
\beqa
z & = & \frac{\mu_n-\mu_p+q_0+Q}{T}.
\eeqa

The above expressions apply for arbitrary values of nucleon degeneracy,
recoil energy, and magnetic field.
In this general case, the nucleon tensor
$N^{\mu\nu}$ depends on $s_n$ and $s_p$, the lepton tensor
$L_{\mu\nu}$ depends on
$N_e$, $p_{e,z}$, and $w_{\perp}\equiv |\bp_{n,\perp}+\bk_{\perp}|$, 
and the response function $S_{s_ns_p}$ depends
on $s_n$, $s_p$, $N_p$, $N_e$, $p_{e,z}$, and $w_{\perp}$. 
To evaluate eq.~(\ref{nicekappaa}), we are left
with two infinite sums and three integrals left to perform.

To make progress, we shall proceed in the next subsection
with an approximate method appropriate to the outer layers of the 
proto-neutron star in which the nucleons are nondegenerate and
the recoil energies and nucleon spin energies are small in comparison
to other energy scales. As discussed before (see \S III), only in the
outer layers (where the neutrino distribution deviates from thermal
equilibrium) can asymmetric neutrino flux develops. 

\subsection{Evaluation of the Absorption Opacity: Nondegenerate Nucleon 
Regime}

For small nucleon spin energies and nucleon recoil, all dependence
on $s_n$, $s_p$, and $N_p$ can be taken out of the exponential in the
nucleon response function so that these quantities can easily be summed
over. Since $S_{s_ns_p}$ is expanded to linear order in $\mu_B B$, 
it contains only terms linear in $s_n$ or $s_p$, but not both $s_ns_p$.
As a consequence, any terms in $L_{\mu\nu}$ containing $s_ns_p$ can
immediately be dropped, considerably simplifying this expression. We
will also drop all terms in $L_{\mu\nu}$ which will give small corrections
to the angle-independent, $B=0$ opacity. Lastly, we drop terms in $L_{\mu
\nu}$ which will give zero in the sums over $N_p$ (see Appendix D for
discussion.)

\subsubsection{ Contribution from the $N_e=0$ State }

Since the electron in the the ground Landau level can only have a
spin opposite the magnetic field, the $N_e=0$ term 
in the opacity expression (eq.~[\ref{nicekappaa}]) requires special 
treatment. As eq.~(\ref{nicekappaa}) already contains a prefactor
$B$, we can drop all nucleon polarization terms when evaluating the $N_e=0$
contribution to the asymmetric opacity (This cannot be done when summing
over all the $N_e \geq 1$ states since $N_e=p_{e,\perp}^2/2eB$ is summed
over a large number of states so that the prefactor of $B$ effectively
cancels). Since only the nucleon polarization terms contain pieces
with large coefficients, the $N_e=0$ state can be evaluated to lowest
order in inelasticity.

In Appendix \ref{absresponseappendix}, it was shown that to lowest order
in the inelasticity, the nucleon response function for $N_e=0$ can be 
written
\beqa 
S_{s_ns_p} & = & \frac{m}{|q_{z,0}|} \exp \left( \frac{\mu_n}{T} 
- \frac{p_{n,\perp}^2}{2mT} - u^2 \right) \ \ \ \ \ 
\mbox{($N_e=0$ state)}, 
\eeqa 
where $u$ is a dimensionless electron z-momentum defined by
\beqa
p_{e,z} & = & \pm (k+Q)(1+\epsilon u),
\eeqa
and 
\beqa
\epsilon & = & \left( \frac{2T}{m} \right)^{1/2} \frac{|q_{z,0}|}{k+Q}
\eeqa
is a small parameter ($q_{z,0}=k_z\mp (k+Q)$). 
As this expression for $S_{s_ns_p}$ is independent of $N_p$, 
we may sum over $N_p$ in $L_{\mu\nu}$, with the result (Appendix D):
\beqa
&& \sum_{N_p=0}^{\infty}N^{\mu\nu}L_{\mu\nu}(N_e=0)  =
\Theta(p_{e,z}) (c_V^2-c_A^2) \bo \cdot \bbhat.
\eeqa
Since this expression is independent of $s_n$ and $s_p$, their sums
give a factor of four. The asymmetric opacity from the electron ground state 
is then given by
\beqa
&&\kabs(N_e=0)  =  \frac{G_F^2}{2} \frac{eB}{2\pi}
\int_{-\infty}^{\infty}  \frac{ dp_{e,z}}{2\pi} (1-f_e)
\int \frac{d^2p_{n,\perp}}{(2\pi)^2}
\nonumber \\ && 
~~~~~~~~~\times 4\frac{m}{|q_{z,0}|} \exp \left( \frac{\mu_n}{T}
- \frac{p_{n,\perp}^2}{2mT} - u^2 \right)
\Theta(p_{e,z}) (c_V^2-c_A^2) \bo \cdot \bbhat
\nonumber \\ &&
~~~~~=\frac{G_F^2}{2} \frac{eB}{2\pi} 4 \frac{mT}{2\pi}
\frac{m}{|q_{z,0}|} \exp\left({\mu_n\over T}\right)
(c_V^2-c_A^2) \bo \cdot \bbhat
\int_{0}^{\infty}  \frac{ dp_{e,z}}{2\pi} (1-f_e) e^{-u^2} 
\nonumber \\ && 
~~~~~=\frac{G_F^2eBn_n}{2\pi}
\left[ 1 - f_e(k+Q) \right](c_V^2-c_A^2)\bo \cdot \bbhat,
\label{kappane0}\eeqa
where we have used eq.~(\ref{chemical}) to 
relate the neutron chemical potential $\mu_n$ to its number density
$n_n$. In evaluating the $p_{e,z}$ integral in eq.~(\ref{kappane0}),
we have approximated $f_e(E_e)$ by $f_e(k+Q)$ since the first 
term in an expansion of $f_e(E_e)$ about this value is odd in $u$ and 
hence gives vanishing contribution and the second order term in the 
expansion is down by a factor $T/m$.

\subsubsection{ Contribution from the $N_e \geq 1$ States }

For electrons in the excited Landau levels ($N_e\ge 1$), the relevant 
matrix element can be written as (see Appendix D):
\be
N^{\mu\nu}L_{\mu\nu} (N_e \geq 1) =  \cT_0+\cT_e+\cT_n,
\ee
with
\beqa
&&\cT_0={1\over 2}(c_V^2 + 3c_A^2)\left[I^2_{N_e-1,N_p}(\omega)
\left( 1 - \frac{p_{e,z}}{|\Lambda|} \right)
+ I^2_{N_e,N_p}(\omega)
\left( 1 + \frac{p_{e,z}}{|\Lambda|} \right)
\right],\\
&&\cT_e=\frac{1}{2} \left[-I^2_{N_e-1,N_p}(\omega)
\left( 1 - \frac{p_{e,z}}{|\Lambda|} \right)
+ I^2_{N_e,N_p}(\omega)
\left( 1 + \frac{p_{e,z}}{|\Lambda|} \right)
\right] (c_V^2-c_A^2) \bo \cdot \bbhat,\\
&&\cT_n=\left[c_A(c_A+c_V)s_n \bo \cdot \bbhat
- c_A (c_A - c_V) s_p \bo \cdot \bbhat\right]\nonumber\\
&&~~~~~~~~\times
\left[I^2_{N_e-1,N_p}(\omega)
\left( 1 - \frac{p_{e,z}}{|\Lambda|} \right)
+ I^2_{N_e,N_p}(\omega)
\left( 1 + \frac{p_{e,z}}{|\Lambda|} \right)\right],
\eeqa
where $|\Lambda|=(p_{e,z}^2+2eBN_e)^{1/2}$, 
$\omega=w_{\perp}^2/(2eB)=(\bp_{n,\perp}+\bk_\perp)^2/(2eB)$, and 
the function $I_{N_e,N_p}$ (The shape of the Landau wave function)
is defined by eq.~(\ref{wavefunc}). Note that in these expressions, 
we have dropped all terms that will give zero contribution to 
the opacity (such as those terms involving $s_ns_p$).  

The $N_e\ge 1$ contribution to the absorption opacity
(eq.~[\ref{nicekappaa}]) can be written as
\beqa
\kabs(N_e \geq 1) & = & \kabs_0 + \kabs(e,N_e \geq 1) + \kabs(np),
\eeqa
where each piece corresponds to a different part of 
$N^{\mu\nu}L_{\mu\nu}(N_e\ge 1)$:
\beqa
&&\kabs_0  =  \frac{G_F^2}{2} \frac{eB}{2\pi}
\sum_{N_e=1}^{\infty}
\int_{-\infty}^{\infty}  \frac{ dp_{e,z}}{2\pi} (1-f_e)
\sum_{N_p=0}^{\infty} \int \frac{d^2p_{n,\perp}}{(2\pi)^2}
\sum_{s_p,s_n} S_{s_n s_p}\cT_0,\label{kappa0}\\
&&\kabs(e,N_e \geq 1) =  \frac{G_F^2}{2} \frac{eB}{2\pi}
\sum_{N_e=1}^{\infty}
\int_{-\infty}^{\infty}  \frac{ dp_{e,z}}{2\pi} (1-f_e)
\sum_{N_p=0}^{\infty} \int \frac{d^2p_{n,\perp}}{(2\pi)^2}
\sum_{s_p,s_n} S_{s_n s_p}\cT_e,\label{kappae}\\
&&\kabs(np)=  \frac{G_F^2}{2} \frac{eB}{2\pi}
\sum_{N_e=1}^{\infty}
\int_{-\infty}^{\infty}  \frac{ dp_{e,z}}{2\pi} (1-f_e)
\sum_{N_p=0}^{\infty} \int \frac{d^2p_{n,\perp}}{(2\pi)^2}
\sum_{s_p,s_n} S_{s_n s_p}\cT_n.
\label{kappanp}
\eeqa

We are going to evaluate the sum $\sum_{N_e}$ by replacing it with
an integral. Such a procedure effectively eliminates any possible
oscillatory behavior of the opacity as a function of energy (see
the beginning of \S V), but is valid when more than a few
electron Landau levels are filled. For infinite nucleon mass,
energy conservation requires $k+Q=(p_{e,z}^2+2eBN_e)^{1/2}$
(neglecting $m_e$). For given $N_e$ and $p_{e,z}$, the nucleon 
recoil energy is of order $|q_z|\sqrt{T/m}$. Thus it is
natural to define a dimensionless recoil energy $u$ via
\beqa
&&2eBN_e = E_{\perp}^2 (1+\epsilon u),
\label{peperp}
\eeqa
where 
\be
E_{\perp}^2=(k+Q)^2-p_{e,z}^2,~~~~~
\epsilon  = \left( \frac{8T}{m} \right)^{1/2} \frac{|q_z|(k+Q)}
{E_{\perp}^2}.
\ee
The nucleon response function (eq.~[\ref{response}]) can be expanded
for small $\epsilon u$, with the result (Appendix E):
\beqa
S_{s_ns_p} & = & \frac{m}{|q_z|}\exp(-y_0)(1-\delta y),
\eeqa
where
\beqa
\exp(-y_0) & \simeq &
\exp \left( \frac{\mu_n}{T} - \frac{p_{n,\perp}^2}{2mT} - u^2 \right)
\left[ 1 + \frac{\epsilon u^3 E_{\perp}^2}{2(k+Q)^2} -
\frac{2 u (k+Q) q^2}{\epsilon m E_{\perp}^2} \right],
\eeqa
and
\beqa
\delta y & = &
- \frac{\mu_{Bn}s_nB}{2T}\left[ 1 - \frac{q_{\perp}^2}{q_{z}^2}
- \frac{\epsilon u m E_{\perp}^2}{(k+Q)q_{z}^2}\left( 1 - \frac{\epsilon u
E_{\perp}^2}{4(k+Q)^2} \right)
\right]
\nonumber \\ & - &
 \frac{\mu_{Bp}s_pB}{2T}\left[ 1 + \frac{q_{\perp}^2}{q_{z}^2}
+ \frac{\epsilon u m E_{\perp}^2}{(k+Q)q_{z}^2}\left( 1 - \frac{\epsilon u
E_{\perp}^2}{4(k+Q)^2} \right)\right].
\eeqa
Similarly, for small $\epsilon u$, the electron Fermi blocking factor can be
expanded to first order in $\epsilon$ as
\beqa
1-f_e(E_e) & \simeq & 1 - f_e(k+Q) - \frac{ \partial f_e(k+Q)}{\partial
E_e} \frac{\epsilon u E_{\perp}^2}{2(k+Q)}
= \left[ 1 - f_e(k+Q) \right] \left[ 1 + \frac{\epsilon u E_{\perp}^2}
{2T(k+Q)} f_e(k+Q) \right]
\eeqa

Now consider $\kabs_0$ in eq.~(\ref{kappa0}). 
Since only $S_{s_ns_p}$ depends on spin, the spin sums $\sum_{s_n,s_p}$
effectively set $\delta y =0$. The factor $\exp(-y_0)$ can be 
evaluated at lowest order. The sum $\sum_{N_p}$ can be 
calculated using the summation rule for $I_{NS}^2$ (Appendix D).
Replacing $\sum_{N_e}$ by $\int dN_e= \int \epsilon E_\perp^2 du/(2eB)$
and integrating over $p_{n,\perp}$ (see Appendix F), we arrive at
\beqa
&&\kabs_0  =  \frac{G_F^2}{2} \frac{eB}{2\pi}
\left( c_V^2 + 3c_A^2 \right) \frac{mT}{2\pi}\, 4 e^{\mu_n/T}
\int_{-(k+Q)}^{(k+Q)}  \frac{ dp_{e,z}}{2\pi}
\frac{m}{|q_z|}
\int_{-\infty}^{\infty} \frac{\epsilon E_{\perp}^2 du}{2eB}
\exp(-u^2) (1-f_e)
\nonumber \\ &&
=\frac{G_F^2}{2} \frac{eB}{2\pi}
\left( c_V^2 + 3c_A^2 \right) \frac{mT}{2\pi}\, 4 e^{\mu_n/T} m
\left( \frac{8T}{m} \right)^{1/2} \frac{(k+Q)}{2eB} \pi^{1/2}
\int_{-(k+Q)}^{(k+Q)}  \frac{ dp_{e,z}}{2\pi} (1-f_e)
\nonumber \\ && 
=\frac{G_F^2}{\pi} (k+Q)^2 n_n (c_V^2+3c_A^2)[1-f_e(k+Q)],
\label{kappa0result}\eeqa
which is exactly the usual $B=0$ opacity. 

Next consider $\kabs(e,N_e\ge 1)$, the ``electron contribution" from the $N_e
\geq 1$ states to the opacity (eq.~[\ref{kappae}]). 
We may evaluate all quantities to lowest order in the inelasticity. The spin
sums and nucleon response function are evaluated as before. Performing
all integrals but $p_{e,z}$ gives
\beqa
&&\kabs(e,N_e \geq 1)  \propto 
\int_{-(k+Q)}^{(k+Q)} dp_{e,z} p_{e,z} = 0.
\eeqa 
So the electron contribution from the higher Landau levels is zero
to lowest order in the inelasticity. The next order correction scales as
$T/m$, and can be neglected. 

Finally, $\kabs(np)$ (eq.~[\ref{kappanp}]) gives the contribution of nucleon
polarizations to the opacity. 
Performing the spin sums, the integral over $p_{n,\perp}$,
and using the results of Appendix D for the sum over $N_p$ yields
\beqa
&&\kabs(np)  = \frac{G_F^2}{2} \frac{eB}{2\pi} 4 \frac{mT}{2\pi}
e^{\mu_n/T} \bo \cdot \bbhat
\int_{-\infty}^{\infty}  \frac{ dp_{e,z}}{2\pi} \frac{m}{|q_z|}
\sum_{N_e=1}^{\infty} (1-f_e) \exp(-u^2)
\nonumber \\ && \times
\left\{
\frac{\mu_{Bn}B}{T} c_A(c_A+c_V) - \frac{\mu_{Bp}B}{T} c_A (c_A - c_V) 
+ \left[ \frac{\mu_{Bn}B}{T} c_A(c_A+c_V) + 
\frac{\mu_{Bp}B}{T} c_A (c_A - c_V) \right]
\right. \nonumber \\ && \times \left.
\left[ \frac{\epsilon^2 u^2 m E_{\perp}^4}{4(k+Q)^3q_z^2}
- \frac{2eBN_e + k_{\perp}^2 + eBp_{e,z}/|\Lambda|}{q_z^2}
\right. \right. \nonumber \\ && \left. \left.
- \frac{\epsilon u m E_{\perp}^2}{(k+Q)q_z^2}
-  \frac{\epsilon^2 u^4 m E_{\perp}^4}{2(k+Q)^3q_z^2}
+ 2u^2 \frac{ q_z^2+2eBN_e + k_{\perp}^2 + eBp_{e,z}/|\Lambda|}{q_z^2}
\right]
\right\}.
\eeqa  
Again changing the sum over $N_e$ into an integral over $u$, expanding
$f_e$, and then performing the trivial $p_{e,z}$ integral gives the final
result
\beqa
&&\kabs(np)  = \frac{G_F^2}{\pi} (k+Q)^2 n_n \bo \cdot \bbhat
\left[ 1 - f_e(k+Q) \right]
\left\{ 2 \frac{\mu_{Bn}B}{T} c_A(c_A+c_V)
\right. \nonumber \\ && \left.
- \frac{T}{(k+Q)} \left[ 1 + \frac{(k+Q)}{T} f_e(k+Q) \right]
\left[ 2 \frac{\mu_{Bn}B}{T} c_A(c_A+c_V)
+ 2 \frac{\mu_{Bp}B}{T} c_A(c_A-c_V)
  \right]\right\}.
\label{nucleonpolarization}
\eeqa

\subsubsection{ Result }
\label{results}

We can summarize our result for the absorption opacity in the following
transparent formula:
\beqa
\kabs &= & \kabs_0
\left( 1 + \epsilon_{\rm abs} \bo \cdot \bbhat \right),
\label{kappaabs}\eeqa
where $\kabs_0$ is the $B=0$ opacity as given by eq.~(\ref{kappa0result}),
and the asymmetry parameter $\epsilon_{\rm abs}$ is given by
\be
\epsilon_{\rm abs}  =  \epsilon_{\rm abs}(e) + \epsilon_{\rm abs}(np),
\label{epscorrect}\ee
with
\beqa
&& \epsilon_{\rm abs}(e)  =  \frac{1}{2} \frac{eB}{(k+Q)^2}
\frac{c_V^2-c_A^2}{c_V^2+3c_A^2}
\label{epselectron} \\ 
&& \epsilon_{\rm abs}(np) = 
2 \frac{c_A(c_A+c_V)}{c_V^2+3c_A^2} \frac{\mu_{Bn} B}{T}
\nonumber \\ &  &
~~~~~~- \frac{T}{(k+Q)}
\left[ 1+\frac{(k+Q)}{T} f_e(k+Q) \right]
\left[2 \frac{c_A(c_A+c_V)}{c_V^2+3c_A^2} \frac{\mu_{Bn} B}{T} 
+ 2 \frac{c_A(c_A-c_V)}{c_V^2+3c_A^2} \frac{\mu_{Bp} B}{T} 
\right].
\label{epscorrect2}\eeqa
Comparing this result with eq.~(\ref{kappawrong}), which was obtained from
a simplified calculation assuming infinite nucleon mass and neglecting Landau
levels for electrons, we see that the simplified calculation gave an 
incorrect result for the asymmetry parameter. Only a neutron polarization
term was included in eq.~(\ref{kappawrong}). The correct expression
for $\epsilon_{\rm abs}$ 
(eqs.~[\ref{epscorrect}]-[\ref{epscorrect2}]) contains 
an electron contribution (arising from from the ground-state Landau level)
and additional contribution from both neutron and proton polarizations
(which arises from our more careful treatment of inelasticity). 
These news terms dominate the asymmetric parameter for neutrino energy
$k/T\simless$ a few. 
It is interesting to note that the particles in the final state of this
reaction ($p$ and $e^-$) contribute to the asymmetry parameter. 
Previous investigators found a contribution only from the initial state 
particles.

\subsection{Moments of the Absorption/Emission Rate}

With the absorption opacity given in previous subsection, it is 
straightforward to calculate the moments of the absorption/emission
rate in the Boltzmann equation. 
The absorption rate is given by eq.~(\ref{absorb2})
with $\delta f_\nu$
given by eq.~(\ref{expandfnu}). Using the opacity 
(\ref{kappaabs}) and the integrals in eq.(\ref{angularintegrals}),
we obtain the zeroth, first and second moments: 
\beqa
\int \frac{d\Omega}{4\pi} 
\left[{\partial f_\nu(\bk)\over\partial t}\right]_{\rm abs} &=&
 -  \kabsstar_0 \left( g + \epsilon_{\rm abs} \bh \cdot \bbhat \right),
\label{abszeromom}\\
\int \frac{d\Omega}{4\pi} \Omega_i 
\left[{\partial f_\nu(\bk)\over\partial t}\right]_{\rm abs} &=&
 -  \kabsstar_0 \left( h_i + \frac{1}{3}\epsilon_{\rm abs} g \hat{B}_i
+ \frac{2}{3} \epsilon_{\rm abs} I_{ij} \hat{B}_j \right), 
\label{absfirstmom}\\
\int \frac{d\Omega}{4\pi} \calp_{ij}
\left[{\partial f_\nu(\bk)\over\partial t}\right]_{\rm abs} &=&
 -  \kabsstar_0 \left[ I_{ij} + 
\frac{3}{10} \epsilon_{\rm abs} \left( \hat{B}_i h_j + \hat{B}_j h_i
- \frac{2}{3} \delta_{ij} \bh \cdot \bbhat\right) \right]
\label{abssecondmom}
\eeqa  
where
\beqa
\kabsstar_0 & = & \kappa^{({\rm abs})}_0
\left[1+\exp \left(\frac{\mu_{\nu}-k}{T}\right)
\right]
\eeqa
(see eq. [\ref{kappastar}]).

\subsection{Absorption Opacity for $\bar\nu_e$}

So far we have been concerned with the opacity for
$\nu_e+n\rightarrow p+e$. The opacity for 
$\bar\nu_e+p\rightarrow n+e^+$ can been obtained by a similar 
calculation. The result can be found from the equations in 
\S \ref{results} by replacing $Q$ by $-Q$, $n_n$ with $n_p$,
the electron distribution function by the positron distribution function 
(which is rather small), and $\mu_{Bn}$ with $-\mu_{Bp}$.
Thus, the absorption opacity for $\bar\nu_e$ is given by
\beqa
\bar\kabs &= & \bar\kabs_0
\left( 1 + \bar\epsilon_{\rm abs} \bo \cdot \bbhat \right),
\eeqa
where $\bar\kabs_0$ is the $B=0$ opacity as given by
\be
\bar\kabs_0=\frac{G_F^2}{\pi} (k-Q)^2 n_p (c_V^2+3c_A^2)[1-f_{e^+}(k-Q)]
\Theta(k-Q),
\label{kabsbar}
\ee
and the asymmetry parameter $\bar\epsilon_{\rm abs}$ is given by
\be
\bar\epsilon_{\rm abs}  = \bar\epsilon_{\rm abs}(e^+) 
+ \bar\epsilon_{\rm abs}(np),
\ee
with
\beqa
&&\bar\epsilon_{\rm abs}(e^+)  =  \frac{1}{2} \frac{eB}{(k-Q)^2}
\frac{c_V^2-c_A^2}{c_V^2+3c_A^2}\\
&& \bar\epsilon_{\rm abs}(np) = 
-2 \frac{c_A(c_A-c_V)}{c_V^2+3c_A^2} \frac{\mu_{Bp} B}{T}
\nonumber \\ &  &
~~~~~~+ \frac{T}{(k-Q)}
\left[ 1+\frac{(k-Q)}{T} f_{e^+}(k-Q) \right]
\left[2 \frac{c_A(c_A-c_V)}{c_V^2+3c_A^2} \frac{\mu_{Bp} B}{T} 
+ 2 \frac{c_A(c_A+c_V)}{c_V^2+3c_A^2} \frac{\mu_{Bn} B}{T} 
\right].
\eeqa
The theta function in eq.(\ref{kabsbar}) comes from the fact that the
reaction is not energetically allowed unless $k \geq Q$ since the 
proton is lighter than the neutron.

\section{The Moment Equations of Neutrino Transport}
\label{combinedmoments}

In the last two sections (\S IV and \S V) we have carried out detailed
calculations of neutrino scattering and absorption in magnetic fields. 
Explicit expressions have been obtained in the nondegenerate 
nucleon regime, which is appropriate for the outer layer the of proto-neutron
star where asymmetric neutrino flux (drift flux) is expected. 
We now use these results to derive the moments of the Boltzmann
transport equation (\ref{boltzmanneqn}). We focus on $\nu_e$ below,
although similar results can also be obtained for other neutrino species. 

The zeroth moment is obtained by integrating eq.~(\ref{boltzmanneqn})
over $d\Omega$. Combining eqs.~(\ref{scatzeromom}) and (\ref{abszeromom}) 
we find
\beqa
\frac{\partial (\fnu^{(0)}+g)}{\partial t} + \grad \cdot \bh
 = && \int \frac{d\Omega}{4\pi} 
\left[{\partial f_\nu(\bk)\over\partial t}\right]_{\rm sc}
+\int \frac{d\Omega}{4\pi} 
\left[{\partial f_\nu(\bk)\over\partial t}\right]_{\rm abs} \nonumber\\
 = && \kscat_0 \frac{k}{m} \Biggl\{
6 \left[ T\frac{\partial g}{\partial k} + (1-2\fnu^{(0)}) g \right]
\nonumber \\ && + 
\frac{k}{T}\left[ T^2 \frac{\partial^2 g}{\partial k^2}
+ T\frac{\partial g}{\partial k}(1-2\fnu^{(0)})
- 2 g T\frac{\partial \fnu^{(0)}}{\partial k} \right] 
\Biggr\}\nonumber \\ && 
+\,\epsilon_{\rm sc} \kscat_0 \left( 
T \frac{\partial \bh}{\partial k} + 4 \frac{T}{k} \bh \right)
\cdot \bbhat-\kabsstar_0 \left( g + \epsilon_{\rm abs} 
\bh \cdot \bbhat \right).
\eeqa
The zeroth moment equation governs the energy exchange between
matter and neutrinos. The $B=0$ part of the scattering opacity can be ignored
in the zeroth moment equation since it is suppressed by a factor of $k/m$
(i.e., only the inelastic part of the scattering contributes to 
matter-neutrino energy exchange; recall that we have not included 
inelastic electron-neutrino scattering which can be a much 
larger effect). The term $-\kabsstar_0 g$ represents the usual 
neutrino emission and absorption. It is of interest to note that 
the asymmetric parts of the scattering and absorption introduce
new terms to the zeroth moment equation. The importance of these
terms will depend strongly on the field strength and optical
depth.
In a steady state (neglecting the time derivative term) we have
\beqa
\grad \cdot \bh
& = & -  \kabsstar_0 g
 -  \epsilon_{\rm abs}\kabsstar_0 \bh \cdot \bbhat 
+ \epsilon_{\rm sc} \kscat_0 \left[ 
T \frac{\partial \bh}{\partial k} + 4 \frac{T}{k} \bh \right] 
\cdot \bbhat.
\label{zero}
\eeqa

The first moment equation is obtained by multiplying
eq.~(\ref{boltzmanneqn}) by $\Omega_i$ and then integrating over $d\Omega$. 
Combining eqs.~(\ref{scatfirstmom}) and (\ref{absfirstmom}) we find
\beqa
&&\frac{\partial h_i}{\partial t} 
+ \frac{1}{3} \frac{ \partial(\fnu^{(0)}+g)}{\partial x_i }
+ \frac{2}{3}\frac{ \partial I_{ij}}{\partial x_j }
 = \int \frac{d\Omega}{4\pi}\Omega_i 
\left[{\partial f_\nu(\bk)\over\partial t}\right]_{\rm sc}
+\int \frac{d\Omega}{4\pi}\Omega_i
\left[{\partial f_\nu(\bk)\over\partial t}\right]_{\rm abs} \nonumber\\
&&~~~~~~~=-\kscat_0 h_i- \frac{1}{3} \epsilon_{\rm sc}\kscat_0 \left\{
\left[ T \frac{\partial g}{\partial k} + ( 1-2\fnu^{(0)})g \right]
\bhat_i+ \left(1 - 2\fnu^{(0)} +\frac{1}{\lambda} - 4 \frac{T}{k}\right)
I_{ij} \bhat_j\right\}\nonumber \\ 
&&~~~~~~~~~~-\kabsstar_0 \left( h_i + \frac{1}{3}\epsilon_{\rm abs} g \hat{B}_i
+ \frac{2}{3} \epsilon_{\rm abs} I_{ij} \hat{B}_j \right).
\eeqa
The  time derivative term can almost always be dropped (This corresponds 
to a rapid redistribution of matter temperature, the timescale of which is
of order the mean free path divided by $c$, much smaller than neutrino
diffusion time of the star; see Ref. \cite{zeldovich}). 
Defining the total $B=0$ opacity, $\ktot_0=\kabsstar+ \kscat_0$, we have
\beqa
h_i =&&
-\frac{1}{3\ktot_0} \frac{ \partial(\fnu^{(0)}+g)}{\partial x_i } 
- \frac{2}{3\ktot_0}\frac{ \partial I_{ij}}{\partial x_j }\nonumber \\ 
&& - \frac{1}{3\ktot_0} \left\{ \epsilon_{\rm sc}\kscat_0
\left[ T \frac{\partial g}{\partial k} + ( 1-2\fnu^{(0)})g \right]
+ \kabsstar_0 \epsilon_{\rm abs} g \right\} \hat{B}_i\nonumber\\
&&- \frac{1}{3\ktot_0} \left[ 2 \kabsstar_0 \epsilon_{\rm abs} + 
\epsilon_{\rm sc}\kscat_0 
\left( 1 - 2\fnu^{(0)} +\frac{1}{\lambda} - 4 \frac{T}{k} \right) 
\right] I_{ij} \hat{B}_j.
\label{one}
\eeqa
Clearly, in addition to the usual diffusive flux (the first line of
eq.~[\ref{one}]), there is also a drift flux (the second and third lines
of eq.~[\ref{one}]) which depends on the direction of the magnetic field. 
This asymmetric drift flux is a unique feature of
parity violation in weak interactions. Equation (\ref{one}) explicitly shows
that the drift flux is nonzero only when the neutrino distribution deviates
from thermal equilibrium, as expected from general consideration of
detailed balance (\S III.C; see also \S II for discussion). The spherical
deviation, $g$, always gives rise to a drift flux along $\bhat$. However,
the drift flux from $I_{ij}$ is along the direction of the vector
$I_{ij}\bhat_j$, which does not have to be directed along the magnetic
field at all points in the star. For cylindrical symmetry, however,
one would expect that the net flux produced by the $I_{ij}\bhat_j$
term would average to the $\bhat$ direction.

Finally, the second moment equation can be obtained by multiplying
eq.~(\ref{boltzmanneqn}) by ${\cal P}_{ij}=(3\Omega_i\Omega_j-\delta_{ij})/2$ 
and then integrating over $d\Omega$. 
Combining eqs.~(\ref{scatsecondmom}) and (\ref{abssecondmom}) and
ignoring the time derivative, we find
\beqa
&& \frac{3}{10} \left( \frac{\partial h_i}{\partial x_j} 
+ \frac{\partial h_j}{\partial x_i}
- \frac{2}{3} \delta_{ij} \grad \cdot \bh \right)
= \int \frac{d\Omega}{4\pi}{\cal P}_{ij}
\left[{\partial f_\nu(\bk)\over\partial t}\right]_{\rm sc}
+\int \frac{d\Omega}{4\pi}{\cal P}_{ij}
\left[{\partial f_\nu(\bk)\over\partial t}\right]_{\rm abs} \nonumber\\
&&~~~~= -\frac{3}{2} \left( \frac{1+3\lambda^2}{1+5\lambda^2} \right)
\kscat_0 I_{ij}- \frac{3}{20} \epsilon_{\rm sc} \kscat_0
\left(1 - 2\fnu^{(0)}+
 \frac{1}{\lambda} - 4 \frac{T}{k} \right)
\left( h_i\bhat_j + h_j\bhat_i -
\frac{2}{3} \delta_{ij} \bh \cdot \bbhat  \right)\nonumber \\ 
&&~~~~~~~  -\kabsstar_0 \left[ I_{ij} +
\frac{3}{10} \epsilon_{\rm abs} \left( \hat{B}_i h_j + \hat{B}_j h_i
- \frac{2}{3} \delta_{ij} \bh \cdot \bbhat\right) \right].
\label{two}
\eeqa

The above equations apply to $\nu_e$. Similar equations can be derived for
other species of neutrinos. Note that since
$\nu_{\mu(\tau)}$ and $\bar\nu_{\mu(\tau)}$ are always created in pairs
inside the proto-neutron star, they have the same energy density 
distribution. Because of the crossing symmetry (see \S IV.F), the
drift flux of $\nu_{\mu(\tau)}$ exactly cancels the drift flux of
$\bar\nu_{\mu(\tau)}$.

\section{Discussion}

In this paper we have presented a detailed study of neutrino-nucleon 
scattering and absorption in strong magnetic fields. Specifically,
we focused on the effect of parity violation in weak interactions 
which can induce asymmetric neutrino transport in the proto-neutron star. 
Starting from the weak interaction Hamiltonian, we found the 
macroscopic moment equations of neutrino transport. Explicit results
applicable to the outer region of a proto-neutron star are given 
in eq.s~(\ref{zero}), (\ref{one}) and (\ref{two}). Despite the fact
that the neutrino cross-sections are asymmetric with respect to the 
magnetic field throughout the star, asymmetric neutrino flux can be
generated only in the outer region of the proto-neutron star where the
neutrino distribution deviates from thermal equilibrium.

Previous studies 
based on simplified treatments (see \S II) have led to misleading
results. We have tried to clarify many of the subtleties in deriving 
the correct expressions. The main technical complication lies 
in the proper treatment of the inelasticity of neutrino-nucleon
scattering/absorption: although these processes are highly elastic 
from the energetics point of view, it is essential to include the
small inelastic effect in order to obtain the correct 
expression for the asymmetric neutrino flux. In addition, it is necessary
to use Landau wave functions for the electron since the 
quantum mechanical ground state of the electron gives the dominant 
contribution to the asymmetry for low energy electron neutrinos.
To obtain simple formulas for the respective opacities, we devoped a 
method to expand phase space integrals for both small
magnetic field strengths (and correspondingly small spin energies) and
also for small inelasticity. This method has general applicability
for computing the effects of nucleon recoil in phase space integrals
in powers of $T/m$.

To quantitatively determine the asymmetry in neutrino 
emission from a magnetized proto-neutron star one has to 
solve the moment equations (\ref{zero}), (\ref{one}) and
(\ref{two}) in the outer layer of the star. This is beyond the
scope of this paper. Here we shall be contented with an 
order-of-magnitude estimate. First note that the net drift flux
associated with $\nu_{\mu(\tau)}$ and $\bar\nu_{\mu(\tau)}$
is zero, and we only need to consider $\nu_e$ and $\bar\nu_e$. 
The key quantities that determine the
asymmetric flux are the dimensionless asymmetry parameters
$\eps_{\rm sc}$ and $\eps_{\rm abs}$. For neutrino-nucleon scattering,
we have (eq.~[\ref{epssc}])
\be
\eps_{\rm sc}\simeq 0.006\,B_{15}T^{-1},
\ee
where $B_{15}$ is the field strength in units of $10^{15}$~G, and $T$ is the
temperature in MeV. The asymmetry parameter for neutrino absorption 
has contributions from electron and nucleons:
\be
\eps_{\rm abs}(e)\simeq 0.6\,B_{15}E_\nu^{-2},~~~~
\eps_{\rm abs}(np)\simeq 0.006\,B_{15}T^{-1}
\left[1+{\cal O}(T/E_\nu)\right], 
\ee
where $E_\nu$ is the neutrino energy in MeV.
Thus for high energy neutrinos the asymmetry is dominated by
$\eps_{\rm sc}$ and $\eps_{\rm abs}(np)$ (arising from nucleon polarization,
$\sim \mu_BB/T$), while for lower
energy neutrinos it is dominated by $\eps_{\rm abs}(e)$ (arising
from electrons in the ground Landau level). The electron
neutrinos decouple from matter near the neutrinosphere, 
where typical density and temperature are $\rho\sim 10^{12}$~g~cm$^{-3}$, 
and $T\sim 3$~MeV. For a mean $\nu_e$ energy of $10$~MeV, 
$\epsilon_{\rm abs}$ is greater than $\epsilon_{\rm sc}$. 
The asymmetry in the $\nu_e$, $\bar\nu_e$ flux is approximately given by the
ratio of the drift flux and the diffusive flux, of order
$\epsilon_{\rm abs} [\kappa_0^{\rm (abs)}/\kappa_0^{\rm (tot)}]$.
Averaging over all neutrino species, we find the total asymmetry
in neutrino flux $\alpha\sim 0.2\epsilon_{\rm abs}$.
To generate a kick of a few hundreds per second would require a
dipole field of order $10^{15}-10^{16}$~G.

Since the asymmetric neutrino flux depends crucially 
on the deviation of the neutrino distribution function from thermal 
equilibrium, it is of interest to consider how the function
$g=f_\nu-f_\nu^{(0)}$ (or $I_{ij}$) scales with the 
depth of the star measured from the
surface. Without a magnetic field, we expect $g$ to decrease exponentially
toward zero below the decoupling layer (which is close to the 
neutrinosphere
for $\nu_e$ and $\bar\nu_e$). In the presence of asymmetric
absorption and scattering opacities, this scaling may be modified. 
Inspecting the zeroth moment equation (\ref{zero}),
we may conclude $g\sim \eps\bh\cdot\bbhat$ in the deep
interior of the star (Recall that in radiative equilibrium one always have
$\nabla\cdot\int \bh\, dk=0$). This effect may increase our estimate 
for the asymmetric flux. We hope to address some of these issues in 
a future paper. 

\acknowledgments

We thank O. Grimsrud, Y.-Z. Qian and I. Wasserman for useful 
discussion. D.L. acknowledges support from the Alfred P. Sloan 
research fellowship.

\appendix

\section{The Matrix Element for $\nu-N$ Scattering \label{matrixelement}}

In the usual case in which the spin projection of the particles is
not ``measured", one can sum the matrix element over the
final spins and average over the initial spins.
However, for spin $1/2$ particles in a magnetic field,
the initial and
final state spin dependence in the matrix element {\it cannot} be
immediately summed over since the nucleon distribution function
and the energy conservation delta function both have
spin dependence of the form $ -s\mu_B B /\kt$.
To obtain the differential cross section one needs to
calculate the matrix element $|M_{ss'}(\bo,\bo')|^2$ for initial (final)
nucleon spin $s$ ($s'$) and initial (final) neutrino direction
$\bo$ ($\bo'$). We neglect the effect of Landau levels of proton
in the $\nu-N$ scattering cross-section (but see Appendix C and D). 

The low energy effective Hamiltonian density for neutral current
scattering of a spin-$1/2$ fermion with a neutrino is given
by (see, e.g., Refs.~\cite{Reddy97,Raffelt96})
\beqa
{\cal H}_{\rm int} & = & \frac{G_F}{\sqrt{2}}
\overl\Psi_N' \gamma_{\mu} \left( \cv - \ca \gamma_5 \right) \Psi_N
\overl\Psi_\nu' \gamma^{\mu} \left( 1 - \gamma_5 \right) \Psi_{\nu}
+ {\mbox h.c.},
\label{hamiltonian}
\eeqa
where neutral current vector and axial coupling constants are 
\cite{Reddy97} given by\footnote{Raffelt and Seckel\cite{Raffelt96,Raffelt95}
considered the isoscalar contributions to the scattering amplitude
as well as the usual isospin pieces, and suggested
$\cv=-1/2$ and $\ca=-1.15/2$ for $\nu + n \rightarrow \nu + n$
and $\cv=1/2-2\sin^2\theta_W$ and $\ca=1.37/2$ for
$\nu + p \rightarrow \nu + p$.}
$\cv=-1/2$ and $\ca=-1.23/2$ for $\nu + n \rightarrow \nu + n$
and $\cv=1/2-2\sin^2\theta_W=0.035$ and $\ca=1.23/2$ for
$\nu + p \rightarrow \nu + p$. Here $G_F=1.166 \times 10^{-5} \gev^{-2}$
is the universal Fermi constant and $\sin^2\theta_W=0.2325$ ($\theta_W$
is the Weinberg angle.
 
The (nonrelativistic) nucleon wavefunction with four momentum
$P=(m+E,\bp)\simeq (m,{\bf 0})$ and spin four-vector $S\simeq s\bhat$  is
given by
\beqa
\Psi_N & = & V^{-1/2} U_N e^{i\bp \cdot \bx - iEt},
\eeqa
(where $V$ is the normalization volume and $U_N$ is the 4-spinor),
while the neutrino wavefunction with four momentum $K=(k,k\bo)$ is
\beqa
\Psi_{\nu} & = & V^{-1/2} U_{\nu}
e^{i\bk \cdot \bx - ikt}.
\eeqa
For the antineutrino, replace $U_\nu \exp \left( i\bk \cdot \bx - ikt 
\right)$ with $V_\nu \exp \left( -i\bk \cdot \bx + ikt \right)$.

Plugging the wavefunctions into eq.~($\ref{hamiltonian}$), the
transition rate $W$ (S-matrix squared divided by time) can be written
\beqa
W \left( \left\{\bp,s,\bo \right\},\left\{\bp',s',\bo' \right\} \right)
& = & \frac{1}{V^3} \left(2 \pi \right)^4 \delta^4
\left( P + K - P' - K' \right)
\left| M_{ss'}(\bo,\bo') \right|^2
\label{microrate}
\eeqa
where
\beqa
\left| M_{ss'}(\bo,\bo') \right|^2 & = & 
\frac{1}{2} G_F^2 L_{\mu \nu} N^{\mu \nu}(s,s'),
\eeqa
\beqa
L_{\mu \nu} & = & 
\overl{U'}_{\nu} \gamma_{\mu}(1-\gamma_5)U_{\nu}
\overl{U}_{\nu} \gamma_{\nu}
(1-\gamma_5)U_{\nu}'
= \frac{1}{4kk'} \tr 
\left[ \Kslash' \gamma_{\mu}(1-\gamma_5)
\Kslash \gamma_{\nu}(1-\gamma_5) \right]
\nonumber \\ &=&   
\frac{2}{kk'}{K'}^{\alpha} K^{\beta} X_{\alpha\mu\beta\nu},
\eeqa
\beqa
X_{\alpha\mu\beta\nu} & = & \frac{1}{4}
 {\rm Tr}\left[ \gamma_{\alpha}\gamma_{\mu}
\gamma_{\beta}\gamma_{\nu}(1-\gamma_5) \right]
=
 g_{\alpha\mu}g_{\beta\nu}
- g_{\alpha\beta}g_{\mu\nu}
+ g_{\alpha\nu}g_{\beta\mu}
- i\epsilon_{\alpha\mu\beta\nu},
\eeqa
\beqa
N^{\mu \nu}(s,s') & = &
\overl{U'}_N \gamma^{\mu}(\cv-\ca \gamma_5)U_N
\overl{U}_N \gamma^{\nu}(\cv-\ca \gamma_5){U'}_N,
\eeqa
and we use the sign conventions $g_{00}=+1$, $g_{ij}=-\delta_{ij}$, and
$\epsilon_{0123}=+1$.
The nucleon piece can be evaluated using
the spin projection operator \cite{bd64} 
$ (1+\gamma_5\gamma_3 s)/2$ 
and the energy projection operator $ (1+\gamma_0)/2$ 
so that
\beqa
&& N^{\mu \nu}(s,s')  =
\frac{1}{4} {\rm Tr} \left[ \left(1+\gamma_5\gamma_3 s' \right) 
\left(1+\gamma_0\right) \gamma^{\mu}(\cv-\ca \gamma_5)
\left(1+\gamma_5\gamma_3 s \right) \left(1+\gamma_0\right)
\gamma^{\nu}(\cv-\ca \gamma_5)\right]
\eeqa
Explicit computation of each component gives
\beqa
N^{00} &= & \frac{1}{2}c_V^2(1+ss')
\nonumber \\
N^{0i} &= & N^{i0}= -\frac{1}{2}c_Vc_A(s+s')\delta_{i3}
\nonumber \\ 
N^{ij} &= & \frac{1}{2}c_A^2 
\left[ \delta_{ij}(1-ss') +2ss'\delta_{i3}\delta_{j3} + 
i\epsilon_{0ij3}(s'-s)\right]
\label{Nmunu}
\eeqa
where $\epsilon_{\mu\nu\lambda\sigma}$ is the completely antisymmetric
tensor with $\epsilon_{0123}=+1$.
The remaining traces can be evaluated by standard methods
\cite{bd64,greiner96} with the result
\beqa
\left| M_{ss'}(\bo,\bo') \right|^2 & = &
\frac{1}{2} G_F^2 c_V^2 \left\{
\left( 1+3 \lambda^2 \right) + \left( 1-\lambda^2 \right) \bo \cdot \bo'
\right. \nonumber \\ && \left.
+ 2\lambda(\lambda+1)(s\bo+s'\bo')\cdot \bbhat
- 2\lambda(\lambda-1)(s\bo'+s'\bo)\cdot \bbhat
\right. \nonumber \\ && \left.
+ ss' \left[ \left( 1-\lambda^2 \right)(1+\bo \cdot \bo')
+4\lambda^2\bo \cdot \bbhat \bo' \cdot \bbhat \right]\right\}
\label{m2}
\eeqa
where we have defined $\lambda=c_A/c_V$.

Time-reversal invariance can be explicitly checked for the matrix 
element in eq. ($\ref{m2}$), or equivalently the S-matrix in 
eq. ($\ref{microrate}$), by simultaneously exchanging all initial
and final state labels.

For antineutrinos, one would just switch $\bo$ and $\bo'$ in eq.
(\ref{m2}).

\section{ Nucleon Response Function for Scattering
\label{nucleonresponse}}

Following the procedure outlined in \cite{Reddy97}, we first
use $d^3p'$ to integrate over $\delta^3(\bp +\bq - \bp' )$ and
then integrate over the azimuthal angle for $\bp$, with the result
\beqa
S_{ss'}(q_0,q) & = & \frac{1}{2\pi}
\int_0^{\infty} dp p^2 \int_{-1}^{1} d\mu 
\delta(q_0+E-E')f_N(E) \left[1 - f_N(E') \right],
\eeqa
where $\mu=\bp \cdot \bq/pq$ and $E' = -\mu_B Bs' + (\bp+\bq)^2/2m =
-\mu_B Bs' + (p^2+q^2+2pq\mu)/2m$. Care must now be taken to
correctly integrate over the energy-conservation delta function with
the coordinate $\mu$; the integral is only nonzero if the
argument of the delta function is zero for $\mu \equiv \mu_0\in(-1,1)$.
We find
\beqa
\mu_0 & = & \frac{ q_0 - q^2/2m - \mu_B B(s-s')}{pq/m}
\eeqa
so that $\mu_0^2 \leq 1$ for
\beqa
p^2 \geq p_{min}^2 = 
\left[ \frac{q_0-q^2/2m-\mu_B B(s-s')}{q/m} \right]^2.
\eeqa
Changing variables from $p$ to $E=-s\mu_B B+p^2/2m$ in the remaining
integral gives
\beqa
S_{ss'}(q_0,q) & = & \frac{m^2}{2\pi q} 
\int_{E_{\rm min}}^{\infty} dE f_N(E)\left[1 - f_N(E+q_0) \right],
\eeqa
where 
\be
E_{\rm min}=-\mu_B Bs + {p_{min}^2\over 2m}
=-\mu_B Bs +{[q_0-q^2/(2m) - \mu_B B(s-s')]^2\over 4\,(q^2/2m)}
\ee
is the minimum energy allowed for the initial state nucleon in order
for energy and momentum conservation to be satisfied given $q_0,q,s$, and
$s'$. The last integral can be done
by first defining the dimensionless variables
$x=(E-\mu_N)/\kt$, $x_{\rm min}=(E_{\rm min}-\mu_N)/\kt$, and $z=q_0/\kt$,
yielding
\beqa
S_{ss'}(q_0,q) & = & \frac{m^2 \kt}{2\pi q}
\int_{x_{\rm min}}^{\infty} dx \left(\frac{1}{e^x+1} \right)
\left( \frac{1}{1+e^{-x-z}} \right)
\nonumber \\ && 
 =  \frac{m^2 \kt}{2\pi q} \frac{1}{1-e^{-z}}
\ln \left[ \frac{1+\exp(-x_{\rm min})}{1+\exp(-x_{\rm min}-z)} \right].
\eeqa
This expression agrees with \cite{Reddy97} for $B=0$
keeping in mind that our definition of $S$ is a factor of two
smaller than theirs.
 
Expanding $x_{\rm min}$ to linear order in $B$ we find
\beqa
x_{\rm min} & \simeq & x_0 + \delta x
\nonumber \\
x_0 & = & \frac{ (q_0-q^2/2m)^2}{ 4\kt(q^2/2m) } - \frac{\mu_N}{\kt}
\nonumber \\
\delta x & = & \frac{-\mu_B B } {2\kt }
\left[ \left( 1 + \frac{2mq_0}{q^2} \right)s +
\left(1 - \frac{2mq_0}{q^2} \right)s' \right].
\label{deltax}
\eeqa
For $\delta x \ll 1$, $S_{ss'}$ can be written as a sum of $S_0$, the 
zero field value, and $\delta S_{ss'}$, the correction due to the 
magnetic field, i.e.,
\beqa
S_{ss'}(q_0,q) & = & S_0(q_0,q) + \delta S_{ss'}(q_0,q)
\nonumber \\
S_0(q_0,q) & = & \frac{m^2 \kt}{2\pi q}
\frac{1}{1-e^{-z}}
\ln \left[ \frac{1+\exp(-x_0)}{1+\exp(-x_0-z)} \right]
\nonumber \\
\delta S_{ss'} & = & - \frac{m^2 \kt}{2\pi q}
\frac{ \delta x}
{ \left[ \exp(x_0)+1 \right] \left[ 1 + \exp(-x_0-z) \right] }.
\eeqa

Note that the asymmetry in the the coefficients in 
eq.~($\ref{Apm}$) is entirely due to the $2mq_0/q^2$ terms, which
first appear in eq.~($\ref{deltax}$) as a consequence of the energy
and momentum conservation delta function. Had one initially set
$m \rightarrow \infty$ in $\delta(E+q_0-E')$, these terms would not
have appeared.  
	
	In the limit of nondegenerate nucleons with 
$\mu_N/T \ll -1$ and $\exp(\mu_N/T) = (2^{1/2}\pi^{3/2}n)/
(m^{3/2}T^{3/2})$, we find
\beqa
&& \frac{1}{1-e^{-z}}
\ln \left[ \frac{1+\exp(-x_0)}{1+\exp(-x_0-z)} \right]
 \simeq  
\frac{ 1}
{ \left[ \exp(x_0)+1 \right] \left[ 1 + \exp(-x_0-z) \right] }
\nonumber \\ 
&& ~~~~~~~~~~~~~\simeq \exp \left( -x_0 \right) = 
\exp\left[ \frac{\mu_N}{T} -
 \frac{ (q_0-q^2/2m)^2}{ 4\kt(q^2/2m) } \right],
\eeqa
which takes the form of a Gaussian in $k'$. The center of the Gaussian
is located at $k'\simeq k$ and has a width
of order $(T/m)^{1/2}k$ due to the recoil motion of the nucleons.
We can further simplify this expression by defining the small quantity
$\epsilon$ and the dimensionless variable $u$ by
\be
\epsilon = \left[{4(1-\mu')T\over m}\right]^{1/2},~~~~~
u =  \frac{k'-k}{\epsilon k}
\ee
so that $k'=k(1+\epsilon u)$. Then the recoil momentum is
\beqa
q^2 & = & 2k^2(1-\mu') \left( 1 + \epsilon u  \right)
+ {\cal O}(\epsilon k)^2
\eeqa
and 
\beqa
 \frac{ (q_0-q^2/2m)^2}{ 4\kt(q^2/2m) } & \simeq &
 \frac{ \epsilon^2 k^2 u^2 \left[ 1 + \frac{2k(1-\mu')}{\epsilon m u}
 \right] }
{ 4T [2k^2(1-\mu')/2m] (1+\epsilon u) } 
\simeq u^2 \left[ 1-\epsilon u  + 2 \frac{k(1-\mu')}{\epsilon mu}\right].
\eeqa
The criteria for the expansion of the $B=0$ part are $T \ll m$ (so
that $\epsilon \ll 1$) and $k \ll (mT)^{1/2}$.
The $B \neq 0$ terms require
the additional assumptions that both 
$\mu_B B/T \ll 1$ and $(\mu_B B/T)(\sqrt{Tm}/k) \ll 1$.
To first order in $\epsilon$ we then find
\beqa
S_0(q_0,q) & = & \frac{\pi^{1/2}n}{\epsilon k} \exp({-u^2}) 
\left[ 1 - \frac{1}{2}\epsilon u + \epsilon u^3
- 2 \frac{k(1-\mu')}{\epsilon m} u 
\right],
\eeqa
and $\delta S_{ss'}=-S_0\delta x$ involves the quantity
\beqa
\frac{2m q_0}{q^2} & \simeq & - \frac{\epsilon m}{k(1-\mu')}
\left( u - \epsilon u^2 \right).
\eeqa
The range of the variable $u$ is from $u_{min}=-1/\epsilon \ll -1$
to $\infty$. Since $\epsilon \ll 1$ we may extend the lower limit
to $-\infty$ with only exponentially small error.


\section{ Wavefunctions for the Absorption Opacity \label{landau}}

The electron wavefunctions in cylindrical coordinates $(\rho,\phi,z)$
are given in \cite{sokolovternov68}. 
In the standard representation for the Dirac matrices \cite{bd64},
these wavefunctions are written as 
\beqa    
\Psi_e & = & L^{-1/2}e^{ip_{e,z}z - iE_et}U_e(\rho,\phi)
\mbox{\ \ \ , \ \ \ } E_e = ( m_e^2 + 2eBN_e +p_{e,z}^2 )^{1/2}
\label{psie}
\eeqa    
where $p_{e,z}$ is the $z$-momentum, and 
$N_e=0,1,2,\cdots$ is the Landau level
index. The 4-spinor $U_e$ is given by 
\beqa
U_e(\rho,\phi) & = &  \left(\frac{1}{2\pi\lambda^2} \right)^{1/2}
e^{i(N_e-R_e)\phi}
\left(   
\begin{array}{l}
C_1 I_{N_e-1,R_e}(t) e^{-i\phi} \\
iC_2 I_{N_e,R_e}(t) \\
C_3 I_{N_e-1,R_e}(t) e^{-i\phi} \\
iC_4 I_{N_e,R_e}(t)
\end{array}
\right)  
\eeqa 
with
\beqa
\begin{array}{l}
C_1  =   \alpha_{+}A_{+} \\
C_2 =\sigma_e \alpha_{-}A_{+} \\
C_3  =   \sigma_e \alpha_{+}A_{-} \\
C_4 = \alpha_{-}A_{-}
\end{array}
&\ \ \ \ \ \ \ \ \ \ \ &
\begin{array}{l}
\alpha_{\pm} = \sqrt{ \frac{1}{2}
\left(1 \pm \sigma_e \frac{p_{e,z}}{|\Lambda|} \right) } \\
A_{\pm} = \sqrt{ \frac{1}{2}
\left(1 \pm \frac{m_e}{E_e} \right) }
\end{array}
\eeqa
\beqa
I_{nr}(t) & = & \left( \frac{r!}{n!} \right)^{1/2}
e^{-t/2} t^{(n-r)/2} L_r^{(n-r)}(t)
\mbox{ \ \ ,  \ \ }
\int_0^{\infty} dt I^2_{nr}(t) =  1.
\label{wavefunc}\eeqa
Here $\lambda=(eB)^{-1/2}$ is the cyclotron radius, 
$t\equiv \rho^2/2\lambda^2$ 
(not to be confused with the symbol for time),
and $V=LA$ is the normalization volume (with length $L$ along 
the $z$-axis and area $A$ in the $x$-$y$ plane; one particle in volume
$V$). The operators which have been simultaneously diagonalized, and their
corresponding eigenvalues, are perpendicular energy
$E_{\perp}=(eB/m_e)(N_e+1/2)$, $z$-momentum $p_{e,z}$, the radius
of the guiding center $R_{gc}=\lambda (2R_e+1)^{1/2}$ ($R_e=0,1,2,\cdots$)
and the ``longitudinal" spin polarization operator $\Lambda=
{\bf \Sigma} \cdot {\bf \Pi}=\sigma_e(p_{e,z}^2 + 2eBN_e)^{1/2}$,
where ${\bf \Pi} = {\bf p} + e {\bf A}$ and $\sigma_e=\pm 1$.
Note that $\Lambda$ reduces to helicity when $B=0$.
The coefficients $C_i$ are found by requiring the spinors to satisfy the
the Dirac equation, the normalization condition, and by diagonalizing
$\Lambda$. The functions $I_{nr}(t)$ are the same as in
\cite{sokolovternov68}, except that we have written them in terms of the
Laguerre polynomials as defined in \cite{abramsteg99}.
 
As the properties of the ground state are crucial to our results,
we now describe them in some detail. The lowest energy state
has $N_e=0$. As the functions $I_{-1,R_e}$ are not well defined, only
the spin projection $\sigma_{e,0} = -{\rm sign}(p_{e,z})$ is allowed for
$N_e=0$; hence the coefficient of $I_{-1,R_e}$ will be zero. To interpret
this restriction, remember that $\sigma_e$ is the projection of the
spin on the vector $\pi$, {\it not} the magnetic field.
However, by expressing ${\bf \pi}$ in terms of raising and lowering
operators one may easily show that
$(\Sigma_x\pi_x + \Sigma_y \pi_y)\Psi(N_e=0)=0$, and
hence $\Lambda \Psi(N_e=0) = \Sigma_z\pi_z \Psi(N_e=0) =
\Sigma_z p_z \Psi(N_e=0) = -p_{e,z} \Psi(N_e=0)$. Hence the $N_e=0$
state with $\sigma_e=\sigma_{e,0}$ has spin opposite to the magnetic
field, as expected of the electron ground state.
Since only one value of $\sigma_e$ is allowed for $N_e=0$, this state is
expected to have a different form for the matrix element than the
$N_e \geq 1$ states.


We assume that the protons are non-relativistic in which case the
wavefunctions can be written as\cite{bd64}
\beqa
\Psi_p & = & L^{-1/2}e^{ip_{p,z}z - iE_pt}U_p(\rho,\phi),
\eeqa
where
\beqa
U_p(\rho,\phi) &= & \left(\frac{1}{2\pi\lambda^2} \right)^{1/2}
e^{i(R_p-N_p)\phi}
\left(
\begin{array}{l}
\delta_{s_p,+1}I_{R_p,N_p}(t) \\
\delta_{s_p,-1}I_{R_p,N_p}(t) \\
0 \\
0
\end{array}
\right)
\eeqa
The energy for the proton is $E_p = (eB/m)(N_p+1/2) -
\mu_{Bp}Bs_p + p_{p,z}^2/2m$. When there is no subscript on the mass, 
we have approximated $m_n \simeq m_p \equiv m$. The spin projection 
along the
magnetic field is $s_p=\pm 1$. The non-relativistic spinors differ
from the relativistic spinors in that they are completely decoupled
from each other. Furthermore, the different components of the spinors
correspond to different energies due to the anomalous magnetic moment
(Recall that in the electron case,
one could arrange the spinors so that the energy
depends only on $N_e$). This complicates the calculation of the matrix
element since the proton distribution function will now depend on spin
(through the energy) and the matrix element cannot be directly summed
over. Lastly, note that the order of $N$ and $R$ was switched going
from the electron to proton case, since the electron has angular momentum
$L_z(e)=N_e-R_e$ and the proton has angular momentum $L_z(p)=R_p-N_p$ 
due to the sign of the charge. The ground state of the proton is the state
with $N_p=0$ and $s_p=+1$. Both $s_p=\pm 1$ are allowed
for $N_p=0$.


The neutron and neutrino wavefunctions are the same as those used for
the scattering calculation (Appendix A). 


 
\section{ The Matrix Element for Absorption }
\label{absmatrixelement}

The S-matrix is given by
\beqa
S_{fi} &=& -i \int d^4x {\cal H}_{int}
\label{sfi}
\eeqa
where the weak interaction, low energy effective Hamiltonian is
(Refs.~\cite{Reddy97,Raffelt96})
\beqa
{\cal H}_{\rm int} & = & \frac{G_F}{\sqrt{2}}
\overl{\Psi}_p \gamma_{\mu} \left( \cv - \ca \gamma_5 \right) \Psi_n
\overl{\Psi}_e \gamma^{\mu} \left( 1 - \gamma_5 \right) \Psi_{\nu}
+ {\mbox h.c.}
\label{abshamiltonian}
\eeqa
where the neutral current coupling
constants for the absorption process are \cite{Raffelt96}
$\cv=1.00$ and $\ca=1.26$ (we shall use the same notation for these
coupling constants for absorption and scattering, even though their
values are different).

Plugging the wavefunctions from Appendix C and
the Hamiltonian in eq. (\ref{abshamiltonian})
we find
\beqa
&& S_{fi} =  -i \frac{G_F}{\sqrt{2}} L^{-1} V^{-1}
2\pi \delta(E_e+E_p-k-E_n-Q) 2\pi \delta(p_{e,z}+p_{p,z}-k_{z}
-p_{n,z})
\nonumber \\ && \times
\int_0^{\infty} d\rho \rho \int_0^{2\pi} d\phi
e^{i{\bf w}_{\perp} \cdot {\bx}_{\perp} } \overl{U}_p(\rho,\phi)
\gamma_{\mu} \left( \cv - \ca \gamma_5 \right) U_n
\overl{U}_e(\rho,\phi) \gamma^{\mu} \left( 1 - \gamma_5 \right) U_{\nu}
\eeqa
where ${\bf w}_{\perp}=(p_{n,x}+k_{x}){\bf e}_x +
(p_{n,y}+k_{y}){\bf e}_y$, and
${\bf x}_{\perp}=x{\bf e}_x + y{\bf e}_y$.
The transition rate (the square of $S_{fi}$ divided by time)
can be written in the form
\beqa
W_{if}^{\rm (abs)}=
\frac{|S_{fi}|^2}{T} & = & L^{-1} V^{-2} (2\pi)^2
\delta(E_e+E_p-k-E_n-Q)\delta(p_{e,z}+p_{p,z}-k_{z}-p_{n,z})
|M|^2
\eeqa
where we have defined
\beqa
|M|^2 & = & \frac{G_F^2}{2}
\left| \int_0^{\infty} d\rho \rho \int_0^{2\pi} d\phi
e^{i{\bf w}_{\perp} \cdot {\bx}_{\perp} } \overl{U}_p(\rho,\phi)
\gamma_{\mu} \left( \cv - \ca \gamma_5 \right) U_n
\overl{U}_e(\rho,\phi) \gamma^{\mu} \left( 1 - \gamma_5 \right) U_{\nu}
 \right|^2.
\eeqa
The integrals over $\rho$ and $\phi$ can be accomplished using 
eqs.~(4.6) and (4.7) in Ref.~\cite{sokolovternov68}, which in our notation give
\beqa
\int_0^{2\pi} \frac{d\phi}{2\pi} e^{i(N_1-R_1)\phi}e^{i(R_2-N_2)\phi}
e^{-i{\bf w}_{\perp} \cdot {\bx}_{\perp} }
& = & J_{(N_1-R_1)-(N_2-R_2)}(w_{\perp}\rho),
\eeqa
where $J_N(z)$ is the $n$-th Bessel function, and
\beqa
\int_0^{\infty}\frac{d\rho \rho}{\lambda^2} I_{N_1R_1}(t)I_{R_2N_2}(t)
J_{(N_1-R_1)-(N_2-R_2)}(\sqrt{2t}\lambda w_{\perp})
& =& (-1)^{N_2-R_2} I_{N_1N_2}(\lambda^2 w^2_{\perp}/2)
I_{R_1R_2}(\lambda^2 w^2_{\perp}/2).
\eeqa
After performing these two integrals, the matrix element $|M|^2$
can be written as
\beqa
|M|^2 & = & \frac{G_F^2}{2}
I_{R_eR_p}^2(\lambda^2 w^2_{\perp}/2)
\left| \overl{\tilde{U}}_p
\gamma_{\mu} \left( \cv - \ca \gamma_5 \right) U_n
\overl{\tilde{U}}_e \gamma^{\mu} \left( 1 - \gamma_5 \right) U_{\nu}
 \right|^2
\eeqa
where we have defined
\beqa
&& \tilde{U}_p =
\left(
\begin{array}{l}
\delta_{s_p,+1} \\
\delta_{s_p,-1}\\
0 \\
0
\end{array}
\right)
{ \mbox \ \ \ \ \ \ \  {\rm and} \ \ \ \ \ \ \ \ }
\tilde{U}_e = \left(
\begin{array}{l}
C_1 I_{N_e-1,N_p}(\lambda^2 w^2_{\perp}/2) \\
iC_2 I_{N_e,N_p}(\lambda^2 w^2_{\perp}/2) \\
C_3 I_{N_e-1,N_p}(\lambda^2 w^2_{\perp}/2)  \\
iC_4 I_{N_e,N_p}(\lambda^2 w^2_{\perp}/2)
\end{array}
\right).
\eeqa
At this point, the summation rule for the $I_{ns}$ (eq.~[2.24] in
Ref.~\cite{sokolovternov68}) may be used to sum over the guiding center
coordinates with the result
\beqa
\sum_{R_e=0}^{R_{max}}\sum_{R_p=0}^{R_{max}} I^2_{R_eR_p}
& = & \sum_{R_e=0}^{R_{max}} 1 = A \frac{eB}{2\pi}.
\eeqa
Since only the matrix element depends on $\sigma_e$, it may be directly
summed over.
As a result, the matrix element may be put into the form
\beqa
\sum_{R_e=0}^{R_{max}}\sum_{R_p=0}^{R_{max}} \sum_{\sigma_e=\pm 1}
c(N_e,\sigma_e)
|M|^2 & = & \frac{G_F^2}{2} A \frac{eB}{2\pi} L_{\mu\nu} N^{\mu\nu}
\eeqa
where the lepton piece is
\beqa
L_{\mu\nu} & = & \sum_{\sigma_e=\pm 1}c(N_e,\sigma_e)
\overl{\tilde{U}}_e \gamma_{\mu} \left( 1 - \gamma_5 \right)U_{\nu}
\overl{U}_{\nu}\gamma_{\nu}\left( 1 - \gamma_5 \right) \tilde{U}_e
\nonumber \\ &=&
 \frac{1}{2k}
\sum_{\sigma_e=\pm 1}c(N_e,\sigma_e)
{\rm Tr} \left[ \tilde{U}_e \overl{\tilde{U}}_e
\gamma_{\mu} \left( 1 - \gamma_5 \right) \Kslash
\gamma_{\nu}\left( 1 - \gamma_5 \right)  \right]
\label{lmunu}
\eeqa
and the nucleon piece is
\beqa
N^{\mu\nu} & = & \overl{\tilde{U}}_p
\gamma^{\mu} \left( \cv - \ca \gamma_5 \right) U_n
\overl{U}_n \gamma^{\nu} \left( \cv - \ca \gamma_5 \right)
\tilde{U}_p.
\eeqa
A moment of inspection shows that the form of the nucleon tensor is
exactly the same as eq.~(\ref{Nmunu}) for the scattering problem
if we replace the
initial nucleon with the neutron and the final nucleon with the proton.
The dependence on the shape of the proton wavefunctions is contained
entirely in the $I_{N_e,N_p}$ functions.

The lepton tensor is more complicated. In particular, notice that
the $N_e=0$ electron Landau level is the only state for which there is
only one polarization. Consequently, the matrix element will have
quite a different structure for the electron ground Landau level and
will contain the important electron contribution to the parity violation
effect. The answers for $N_e=0$ and $N_e\ge 1$ 
will be given separately.

In the $N_e=0$ case, by representing $\tilde{U}_e \overl{\tilde{U}}_e$
in terms of gamma matrices, performing the traces, and then summing
against $N^{\mu \nu}$, we find
\beqa
&& L_{\mu\nu}N^{\mu\nu}(N_e=0)  =  
\Theta(p_{e,z}) I_{0,N_p}^2(w_{\perp}^2/2eB)
\left[ c_V^2 + 3c_A^2 + (c_V^2-c_A^2) \bo \cdot \bbhat 
\right. \nonumber \\ && \left.
+  2c_A(c_A+c_V)(s_p + s_n \bo \cdot \bbhat)
- 2c_A (c_A - c_V)(s_n + s_p \bo \cdot \bbhat)
\right. \nonumber \\ && \left. 
+  s_ns_p \left\{ c_V^2-c_A^2 + (c_V^2+3c_A^2)\bo \cdot \bbhat \right\}
\right]
\nonumber \\ & & \rightarrow
\Theta(p_{e,z}) I_{0,N_p}^2(w_{\perp}^2/2eB) 
(c_V^2-c_A^2) \bo \cdot \bbhat,
\eeqa
where we have taken the relativistic limit ($m_e\rightarrow 0$)
for the electrons. For simplicity, several
terms in this expression which do not affect the final result for the
opacity have been discarded.
First, we have thrown away terms which will only give corrections 
to the angle-independent piece of the opacity. Second,
any terms with $s_ns_p$ have been dropped since they will always give zero in
the sum over $s_n$ and $s_p$ (since we are expanding the nucleon response
function to linear order in the spin energies). Finally, we have
dropped terms like $s_n \bo \cdot \bbhat$ and $s_p \bo \cdot \bbhat$ since 
the $N_e=0$ contribution is already proportional to $B$; these terms will 
yield an additional factor of $B$ from the nucleon polarization which 
will be much smaller.

We shall need $L_{\mu\nu}N^{\mu\nu}(N_e=0)$ summed over all $N_p$. Using the
summation rule for $I_{NS}$:
\beqa
\sum_{S=0}^{\infty} I_{NS}^2 (x) & = & 1, 
\label{sumrule1}
\eeqa
we find
\beqa
&& \sum_{N_p=0}^{\infty}L_{\mu\nu}N^{\mu\nu}(N_e=0)  =  
\Theta(p_{e,z}) (c_V^2-c_A^2) \bo \cdot \bbhat,
\eeqa
where $\Theta$ is the step function.

For the $N_e \geq 1$ case, all terms containing a factor $s_ns_p$,
which gives zero in the sums over $s_n$ and $s_p$, 
and corrections to the angle independent opacity will be dropped.
Furthermore, we drop terms with no spin dependence which are proportional
to $I_{N_e,N_p}I_{N_e-1,N_p}$. Since these terms
have no spin dependence, they
cannot couple to the nucleon polarization terms in the response function
(see Appendix E),
and hence they can be evaluated to lowest order in the inelasticity. In
the sum over $N_p$, $I_{N_e,N_p}I_{N_e-1,N_p}$ will then give zero.
For relativistic electrons we then find
\beqa
&& L_{\mu\nu} N^{\mu\nu} (N_e \geq 1) =  \frac{1}{2} \left[ 
I^2_{N_e-1,N_p}(w_{\perp}^2/2eB) 
\left( 1 - \frac{p_{e,z}}{|\Lambda|} \right)
+ I^2_{N_e,N_p}(w_{\perp}^2/2eB) 
\left( 1 + \frac{p_{e,z}}{|\Lambda|} \right)
\right]
\nonumber \\ && ~~~\times
\left[ c_V^2 + 3c_A^2 +  2c_A(c_A+c_V)s_n \bo \cdot \bbhat 
- 2c_A (c_A - c_V) s_p \bo \cdot \bbhat
\right]
\nonumber \\ &&
~~~+ \frac{1}{2} \left[
- I^2_{N_e-1,N_p}(w_{\perp}^2/2eB) 
\left( 1 - \frac{p_{e,z}}{|\Lambda|} \right)
+ I^2_{N_e,N_p}(w_{\perp}^2/2eB)  
\left( 1 + \frac{p_{e,z}}{|\Lambda|} \right)
\right] (c_V^2-c_A^2) \bo \cdot \bbhat.
\eeqa
Summing this expression over all $N_p$ gives
\beqa
&& \sum_{N_p=0}^{\infty} L_{\mu\nu} N^{\mu\nu} (N_e \geq 1) = 
c_V^2 + 3c_A^2 +  2c_A(c_A+c_V)s_n \bo \cdot \bbhat
\nonumber \\ &&
~~~~~- 2c_A (c_A - c_V) s_p \bo \cdot \bbhat
+  \frac{p_{e,z}}{|\Lambda|} (c_V^2-c_A^2) \bo \cdot \bbhat.
\eeqa
In addition, it will be necessary to sum the matrix element against
nucleon recoil terms from the response function which contains
\beqa
q_{\perp}^2 & = & eB(2N_p+1) - p_{n,\perp}^2.
\eeqa
The needed summation rule is \cite{sokolovternov68}
\beqa
\sum_{S=0}^{\infty} S I^2_{NS}(x) & = & N + x.
\eeqa
In our case this gives
\beqa
\sum_{N_p=0}^{\infty} N_p I^2_{N_eN_p}(w_{\perp}^2/2eB) & = & N_e + 
w_{\perp}^2/2eB, 
\eeqa
so that
\beqa
\sum_{N_p=0}^{\infty} q_{\perp}^2 I^2_{N_eN_p}(w_{\perp}^2/2eB)
& = & 2eB\left( N_e + w_{\perp}^2/2eB \right) + eB - p_{n,\perp}^2
\nonumber \\ 
&= & eB(2N_e+1) + k_{\perp}^2 +   2 \bk_{\perp} \cdot \bp_{n,\perp}.
\eeqa
Note that the large neutron momentum terms ($p_{n,\perp}^2$)
cancelled so that the ``averaged" recoil momentum has the expected size. The
final result needed is then
\beqa
&& \sum_{N_p=0}^{\infty} q_{\perp}^2 L_{\mu\nu} N^{\mu\nu} (N_e \geq 1)
= \left[ 2eBN_e + k_{\perp}^2 +   2 \bk_{\perp} \cdot \bp_{n,\perp}
+ eB \frac{p_{e,z}}{|\Lambda|} \right]\nonumber\\
&&~~~~\times\left[ c_V^2 + 3c_A^2 +
 2c_A(c_A+c_V)s_n \bo \cdot \bbhat
- 2c_A (c_A - c_V) s_p \bo \cdot \bbhat\right]\nonumber \\ 
&&~~~~+\left[ eB + \left( 2eBN_e + k_{\perp}^2 + 
2 \bk_{\perp} \cdot \bp_{n,\perp} \right) \frac{p_{e,z}}{|\Lambda|} 
\right](c_V^2-c_A^2) \bo \cdot \bbhat.
\eeqa


\section{ Nucleon Response Function for Absorption}
\label{absresponseappendix}

The response function for absorption is defined 
in eq.~(\ref{absresponse}).
Using $p_{p,z}$ to integrate over the $z$-momentum delta function 
gives $p_{p,z}=p_{n,z}+k_z-p_{e,z}$. Using $p_{n,z}$ to integrate
over the energy delta function gives
\beqa
p_{n,z} &= & \frac{q_0+Q-q^2/2m + ( \mu_{Bp}s_p-\mu_{Bn}s_n)B}{q_z/m},
\eeqa
where we have defined the energy transfer by $q_0\equiv k-E_e$ 
and the momentum transfer by $q_z\equiv k_z-p_{e,z}$, $q_{\perp}^2
\equiv eB(2N_p+1)-p_{n,\perp}^2$, and $q^2\equiv 
q_\perp^2+q_z^2$. The result is 
\beqa
S_{s_ns_p} & = & \frac{m}{|q_z|} f_n(E_n)\left[1-f_p(E_p)\right],
\eeqa
where the neutron energy is
\beqa
E_n &= & \frac{ p_{n,\perp}^2}{2m} + \frac{ p_{n,z}^2}{2m}-\mu_{Bn}Bs_n
= \frac{p_{n,\perp}^2}{2m} + 
\frac{ \left[q_0+Q-q^2/2m + 
( \mu_{Bp}s_p-\mu_{Bn}s_n)B \right]^2}{4(q_z^2/2m)}
- \mu_{Bn}s_nB, 
\eeqa
and the proton energy is $E_p=Q+E_n+q_0$. 
Defining dimensionless parameters $y,~z$ via
\beqa
y & \equiv & \frac{E_n-\mu_n}{T} = - \frac{\mu_n}{T}
+ \frac{ p_{n,\perp}^2}{2mT} +
\frac{ \left[q_0+Q-q^2/2m +
( \mu_{Bp}s_p-\mu_{Bn}s_n)B \right]^2}{4T(q_z^2/2m)}
- \frac{\mu_{Bn}s_nB}{T},
\eeqa
and 
\beqa
z & = & \frac{\mu_n-\mu_p+q_0+Q}{T},
\eeqa
the response function can be written as
\beqa
S_{s_ns_p} & = & \frac{m}{|q_z|} \left( \frac{1}{e^y+1} \right)
\left( \frac{1}{1+e^{-y-z}} \right).
\eeqa

The above expression is exact. We now consider the regime where
the nucleons are nondegenerate. This is valid for the outer layers of
the neutron star where asymmetric flux can develop.  
Since the nucleon spin energies are small, 
we shall expand $S_{s_ns_p}$ to first order in $\mu_B B$.
Using the nondegenerate nucleon conditions $f_p\ll 1$ and $e^{-y}\ll 1$,
we find:
\beqa
S_{s_ns_p} & = & \frac{m}{|q_z|} e^{-y_0}\left( 1 - \delta y \right),
\label{sabs}
\eeqa
where $y \simeq y_0+\delta y$, and 
\beqa
y_0 & = & - \frac{\mu_n}{T} + \frac{ p_{n,\perp}^2}{2mT}
+ \frac{ (q_0+Q-q^2/2m)^2 }{4T(q_z^2/2m)}
 \nonumber \\
\delta y & = & 
- \frac{\mu_{Bn}s_nB}{2T}\left( 1 + \frac{q_0+Q-q_{\perp}^2/2m}{q_z^2/2m}
\right)- \frac{\mu_{Bp}s_pB}{2T}
\left( 1 - \frac{q_0+Q-q_{\perp}^2/2m}{q_z^2/2m}\right).
\label{y0deltay}
\eeqa
The $\delta y$ term now contains all the dependence on the nucleon spins.
This term will give rise to the nucleon contribution to the asymmetric
parity violation effect.

In evaluating the absorption opacity (eq.~\ref{nicekappaa}), 
it will be necessary to expand $S_{s_n s_p}$ for small ``inelasticity". 
If the nucleon mass were infinite, energy conservation would give
exactly $E_e=k+Q$. Thus we expect $S_{s_ns_p}$ to be sharply peaked
about this electron energy, with a width proportional to $(T/m)^{1/2}$.
We can expand the electron energy around the peak in a series in the 
small parameter $(T/m)^{1/2}$. There are two cases
to consider: for the electron ground state ($N_e=0$), we shall want to
define the dimensionless electron energy in terms of $p_{e,z}$;
but for the case in which we are summing over a continuum of electron
Landau levels it will be more convenient to define the dimensionless
electron energy in terms of the perpendicular momentum $p_{e,\perp}^2
\equiv 2eBN_e$.

In the $N_e=0$ case, $E_e\simeq |p_{e,z}|$ (neglecting $m_e$),
we define the dimensionless electron energy $u$ by
\beqa
p_{e,z} & = & \pm (k+Q)(1+\epsilon u),
\eeqa
where
\beqa
\epsilon & = & \left( \frac{2T}{m} \right)^{1/2} \frac{|q_{z,0}|}{k+Q},
~~~~~q_{z,0}=k_z\mp (k+Q).
\eeqa
(since $\epsilon \ll 1$, we can set $|1+\eps u|=1+\eps u$
over the interesting range of $u$).
As discussed in \S V.B, we will only need the dominant 
term for the $N_e=0$ response function. 
Thus we can drop the nucleon polarization terms in $\delta y$
and work to the lowest order in inelasticity.
With these approximations, we find
\beqa
S_{s_ns_p} & = & \frac{m}{|q_{z,0}|} \exp \left( \frac{\mu_n}{T} 
- \frac{p_{n,\perp}^2}{2mT} - u^2 \right) \ \ \ \ \ 
\mbox{($N_e=0$ state)},
\eeqa
which takes the form of a simple Gaussian in $u$. Since $\epsilon \ll 1$,
we can consider the range of $u$ to extend from $-\infty$ to
$\infty$ in the phase space integrals with exponentially small error.

When summing over $N_e$ in eq.~(\ref{nicekappaa}), 
it will be more convenient to define $u$ in terms of $p_{e,\perp}^2
=2eBN_e$. Let
\beqa
p_{e,\perp}^2 & = & 2eBN_e = E_{\perp}^2 (1+\epsilon u),
\eeqa
where 
\be
E_{\perp}^2=(k+Q)^2-p_{e,z}^2,~~~~~
\epsilon  =\left( \frac{8T}{m} \right)^{1/2} \frac{|q_{z}|(k+Q)}
{E_{\perp}^2}.
\ee
The allowed range of $p_{e,z}$ is now $p_{e,z} \in [-(k+Q),(k+Q)]$ in
order to keep $E_{\perp}^2$ a positive number. The electron energy is
\beqa
E_e & \simeq & \left( p_{e,\perp}^2 + p_{e,z}^2 \right)^{1/2} 
\simeq (k+Q) \left[ 1 + \frac{\epsilon u E_{\perp}^2}{2(k+Q)^2} 
- \frac{\epsilon^2 u^2 E_{\perp}^4}{8(k+Q)^4} \right].
\eeqa
The expressions needed for eq.~(\ref{sabs}) are then
\beqa
\exp(-y_0) & \simeq & 
\exp \left( \frac{\mu_n}{T} - \frac{p_{n,\perp}^2}{2mT} - u^2 \right)
\left[ 1 + \frac{\epsilon u^3 E_{\perp}^2}{2(k+Q)^2} -
\frac{2 u (k+Q) q^2}{\epsilon m E_{\perp}^2} \right],
\eeqa
and 
\beqa
\delta y & = &
- \frac{\mu_{Bn}s_nB}{2T}\left[ 1 - \frac{q_{\perp}^2}{q_{z}^2}
- \frac{\epsilon u m E_{\perp}^2}{(k+Q)q_{z}^2}\left( 1 - \frac{\epsilon u
E_{\perp}^2}{4(k+Q)^2} \right) \right]
\nonumber \\ & - & 
 \frac{\mu_{Bp}s_pB}{2T}\left[ 1 + \frac{q_{\perp}^2}{q_{z}^2}
+ \frac{\epsilon u m E_{\perp}^2}{(k+Q)q_{z}^2}\left( 1 - \frac{\epsilon u
E_{\perp}^2}{4(k+Q)^2} \right) \right]. 
\eeqa 
Note that, as in the scattering case, there is a term in $\delta y$
with a coefficient which scales as
\beqa
\frac{\epsilon m E_{\perp}^2}{(k+Q)q_{z}^2} & \simeq &
\frac{ (mT)^{1/2}}{|q_z|} \sim \left( \frac{ m}{T} \right)^{1/2}
\eeqa
for $k \sim T$, and this term can be much greater than unity. 

\section{ Replacing the Sums over $N_e$ with Integrals}

In this Appendix, we turn the sum over $N_e$ in eq. (\ref{nicekappaa})
into an integral.
Let the sum be called
\beqa
{\cal I} & = & \sum_{N_e=1}^{\infty} F(N_e)
\eeqa
where the function $F$ is given by
\beqa
F(N_e) & = & 
(1-f_e)
\sum_{N_p=0}^{\infty} \int \frac{d^2p_{n,\perp}}{(2\pi)^2}
\sum_{s_n,s_p=\pm 1} S_{s_n s_p}
L_{\mu \nu} N^{\mu \nu}.
\eeqa
Integrating the identity (see Ref.\cite{landaulifshitz})
\beqa
\sum_{n=-\infty}^{\infty} \delta(x-n) & = & \sum_{k=-\infty}^{\infty}
\exp( 2\pi i k x)
\eeqa
against $F(x)$ over the region 
$x \in [1,\infty)$ gives the result
\beqa
\sum_{N_e=1}^{\infty} F(N_e) & = & - \frac{1}{2} F(1) + 
\int_1^{\infty} dN_e F(N_e)\ \  + {\mbox \ \ {\rm oscillatory\ terms} }.
\eeqa 
As discussed at the beginning of \S \ref{absorption}, we shall ignore
the oscillatory terms. 

The $F(1)$ terms will give expressions smaller than the integral over
$N_e$ for both eq. (\ref{kappa0}) and (\ref{kappanp}). The reason is
that the integral over $N_e$ effectively divides by $eB$, so that the
factor of $eB$ is eq. (\ref{nicekappaa}) is cancelled.
For eq. (\ref{kappae}), the $F(1)$ term is odd in $p_{e,z}$ so that
it integrates to zero (to lowest order in inelasticity) in the $p_{e,z}$
integral. The corrections involving inelasticity will make this term
smaller by a factor of $T/m$ than the $N_e=0$ term.
Our result is then
\beqa
{\cal I} & = & 
\int_1^{\infty} dN_e F(N_e).
\eeqa

When changing the variable of integration from $N_e$ to $u$ using the
expression $2eBN_e=E^2_{\perp}(1+\epsilon u)$ in eq.(\ref{peperp}), 
we can let $u$ range from $-\infty$ to $\infty$ since $\epsilon \sim
(T/m)^{1/2} \ll 1$.


\end{document}